\documentclass[11pt]{article}

\usepackage{jheppub}

\usepackage{amsmath,amsfonts,amsthm,graphicx}

\allowdisplaybreaks[3]
\numberwithin{equation}{section}

\usepackage[dvipsnames]{xcolor}

\def\eps{{\epsilon}}

\def\p{\partial}

\def\XXint#1#2#3{{\setbox0=\hbox{$#1{#2#3}{\int}$ }
\vcenter{\hbox{$#2#3$ }}\kern-.6\wd0}}

\def\<{\langle}
\def\>{\rangle}

\def\i2{\frac{i}{2}}

\def\tr{{\rm tr~}}

\def\sH{{\slash\!\!\!\! H}}
\def\sF{{\slash\!\!\!\! F}}

\usepackage{amssymb}
\usepackage{amsbsy}
\usepackage{amsmath}
\usepackage[utf8]{inputenc}

\newcommand{\cW}{\mathcal{W}}
\newcommand{\bF}{\mathbf F}

\newcommand{\bG}{\mathbf G}

\newcommand{\dn}{{\rm dn}\!~}
\newcommand{\cn}{{\rm cn}\!~}
\newcommand{\sn}{{\rm sn}\!~}

    \newcommand{\beq}{\begin{equation}}
    \newcommand{\eeq}{\end{equation}}
    \newcommand\beqa{\begin{eqnarray}}
    \newcommand\eeqa{\end{eqnarray}}

\def\<{\left<}
\def\>{\right>}
\def\d{\partial}

\newcommand{\nn}{\nonumber}
\newcommand{\eq}[1]{(\ref{#1})}

\title{Disc partition function of 2d $R^2$ gravity \\ from DWG matrix model}

\emailAdd{kazakov$\bullet$lpt.ens.fr}
\emailAdd{fedor.levkovich$\bullet$gmail.com}

\author[a,b]{Vladimir Kazakov,}
\author[c,1]{Fedor Levkovich-Maslyuk\note{Also at Institute for Information Transmission Problems, Moscow 127994, Russia}}

\affiliation[a]{
Laboratoire de Physique de l’\'{E}cole Normale Sup\'{e}rieure,
CNRS, Universit\'{e} PSL, Sorbonne Universit\'{e}s,
24 rue Lhomond, 75005 Paris, France
} 

\affiliation[b]{Interdisciplinary Scientific Center Poncelet (CNRS UMI 2615), 119002 Moscow, Russia}

\affiliation[c]{
Universit\'{e} Paris Saclay, CNRS,  CEA, Institut de physique th\'{e}orique,   91191, Gif-sur-Yvette, France
}

\abstract{We compute the sum over flat surfaces of disc topology  with arbitrary number of conical singularities.  To that end, we explore and generalize a specific case of the matrix model of dually weighted graphs (DWG) proposed and solved by one of the authors,  M.~Staudacher and Th.~Wynter. Namely, we compute the sum over quadrangulations of the disc with certain boundary conditions, with parameters  controlling the number of squares (area), the length of the boundary and the coordination numbers of vertices.  The vertices introduce conical defects with angle deficit given by a multiple  of \(\pi\), corresponding to positive, zero  or negative  curvature.  Our results interpolate between the well-known 2d quantum gravity solution for the disc with fluctuating 2d metric  and  the  regime of ``almost flat"  surfaces with all the negative curvature concentrated  on the boundary. We also speculate on possible ways to study the fluctuating 2d geometry with \(AdS_2\) background instead of the flat one.} 

\begin{document}

\maketitle

\section{Introduction}

Matrix integrals at large size \(N\) of the matrices are a formidable tool for counting various types of planar graphs, i.e. ``fat" graphs which have a certain two dimensional topology~\cite{tHooft:1973alw,Brezin:1977sv,Itzykson:1979fi,Migdal:1983qrz}.
 The topological property of the \(1/N\) expansion  in (multi)-matrix integrals and in matrix field theories, applied to  Feynman graphs of asymptotically large sizes, is  the key element for the study of  quantum fluctuations of \(2d\)  geometries  -- \(2d\)  quantum gravity --  with various embedding, as well as the statistical mechanics of spins on planar graphs~\cite{David:1984tx,Kazakov:1985ds,Kazakov:1985ea,Kazakov:1986hy,Kazakov:1988ch,Kazakov:1989bc,Kostov:1988fy,Kostov:1999qx,Kostov:1997bn,Eynard:2007kz}. The famous AdS/CFT  correspondence~\cite{Maldacena:1997re,Witten:1998qj,Gubser:1998bc} is also based on the deep relation between planar graphs and string worldsheets.   More recently, the topological expansion in a specific one-matrix model  was proposed for the description of Jackiw-Teitelboim (JT) gravity~\cite{Saad:2019lba,Witten:2020wvy,Turiaci:2020fjj,Maxfield:2020ale,Mertens:2020hbs,Johnson:2019eik}.    

Usually, matrix models and field theories have couplings in their actions which control the numbers of vertices of certain types in the corresponding planar graphs, but not the numbers of faces of given types.  For instance, planar Feynman graphs in the matrix scalar field theory with \(\lambda \tr\phi^4\) interaction  have weights \(\lambda^n\), where \(n\) is the number of quartic vertices. But the numbers of faces of given size \(n^*=1,2,3\cdots\) are arbitrary (up to the constraint imposed by the Euler theorem) and not weighted by independent couplings. 

\begin{figure}[h]   
\begin{center}
\includegraphics[scale=0.6]{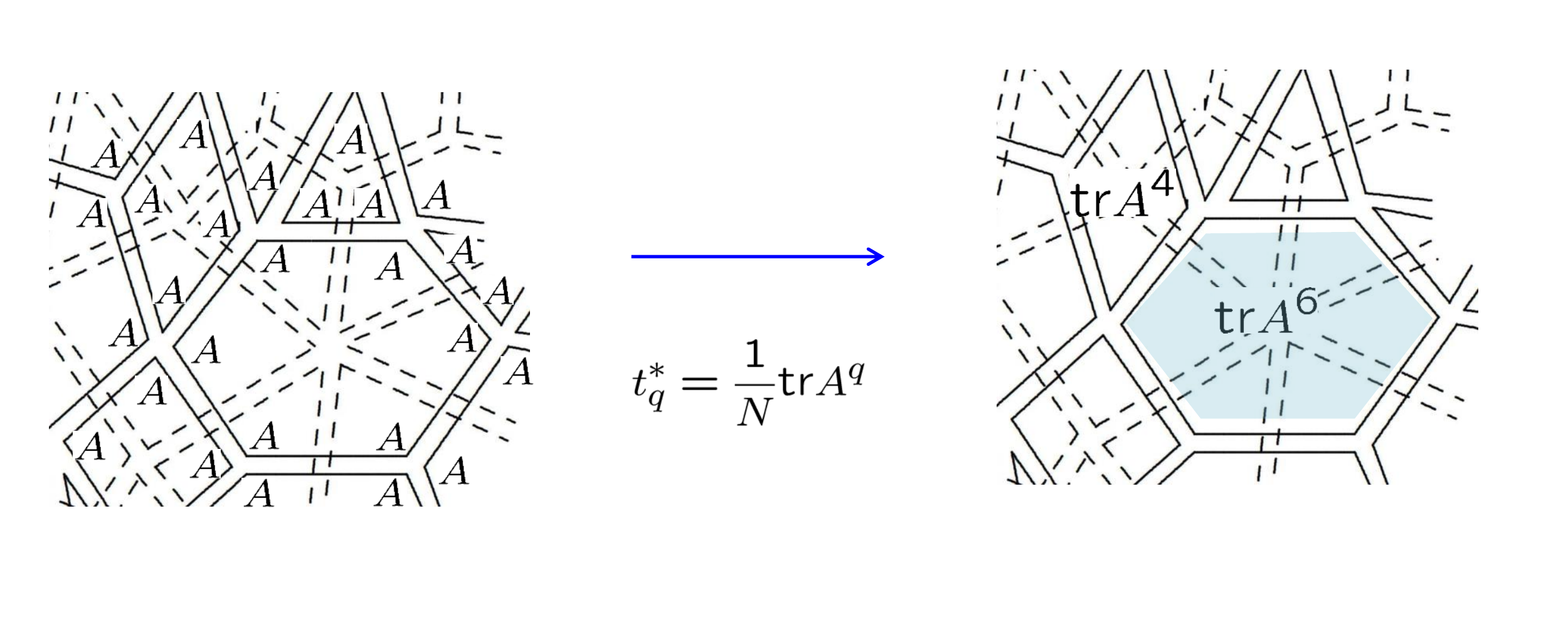}
\end{center}
\caption{A dually weighted Feynman graph. Solid double lines show the propagators, while the dashed lines show the dual graph (whose vertices correspond to faces of the original graph and vice versa). At each $n$-vertex we have $n$ copies of the matrix $A$ which after contraction give factors $\propto\tr A^q$ corresponding to each $q$-face.}
\label{fig:facegr}
\end{figure}

There exists an elegant way to modify a matrix quantum field theory so that also the faces of the Feynman graphs (vertices of the dual graphs) will be weighted with dual couplings attached to the faces of a given order\footnote{and factors depending on the type of interactions }. For example, for the scalar theory one can modify to that end the action as follows: 
\begin{equation}\label{Phi4A}
 Z(t,t^*)=
\int {\cal D}^{N^2}\phi(x)\,\exp N \tr \left(-\frac{1}{2}(\nabla\phi)^2-\frac{m^2}{2}\phi^2+\sum_{q=1}^{Q+1} \frac{1}{q}t_q (A\phi)^q\right)\, \ ,
\end{equation}
where we define the dual couplings as \( t_n^*=\frac{1}{N}\tr A^{n}\).
Then the partition function is given by a perturbative expansion over planar dually weighted Feynman graphs (DWG)  
 \begin{equation}\label{QFTDWG}
\log Z(t,t^*)=
\sum_{G} N^{2-2g_G}\,\, \prod_{v^*_{q_{*}},v_q \in G} w_{G}(\# v^*_{q_*},
\# v_{q}){\,\,\,t_{q_*}}^{\# v^*_{q_*}}\ {t_{q}}^{\# v_{q}}\ ,\end{equation}
where the sum goes over all planar Feynman graphs \(G \),   \(g_G\) is the genus of the graph,  $v_{q_*}^*$ and $v_q$ are the numbers of vertices with $q^*\) and \(q$ neighbours on the
dual graph
and the original graph, respectively, while \(\# v_{q_*}^*,\# v_q\) are the numbers
of such vertices in a given graph. The weight \(w_{G}(\# v^*_{q_*},\# v_{q})\) is the contribution of graphs with fixed  \(\# v^*_{q_*},\# v_{q}\)  and \(g_{G}\) (the result of computation of the corresponding Feynman integrals). This identification of dual couplings and the expansion \eqref{QFTDWG} can be derived from the simple fact that the vertices  in the planar graph expansion corresponding to  the terms  $t_q\tr(A\phi)^q$ in the action   can be  presented as shown on figure~\ref{fig:facegr}.    It is clear that this  results in contracting the product of \(q_*\)  matrices \(A\) in the index loop around each face  with \(q_*\) edges, giving the factor \(t_q^*=\frac{1}{N}\tr A^{q}\), in the normalization  appropriate for the 't~Hooft limit.~\footnote{Curiously, this idea of introducing dual couplings in matrix models  can be used for constructing a quantum field theory in any dimension at finite $N_c$ as a  zero-dimensional matrix model in a specific large $N$ limit~\cite{Kazakov:2000ar} ($N_c$ and $N$ here are
different parameters).}

In this work, we will continue the study of DWG and the matrix model formulated in various forms 
 in~\cite{Das:1989fq,DiFrancesco:1992cn,Kazakov:1995ae}  and solved in the planar limit  in the series of works~\cite{Kazakov:1995ae,Kazakov:1995gm,Kazakov:1996zm,Kazakov:1996vy}. It is the zero-dimensional version of matrix QFT~\eqref{Phi4A}. It represents a natural generalization of the one-matrix model with a general potential, like in \cite{Brezin:1977sv,Kazakov:1989bc} where the  introduction of the \(A \) matrix in the interaction potential  allows one to control not only the order of the vertices of planar graphs (coordination number, or the number of nearest neighbors), but also the order of their faces (coordination numbers of vertices of the dual graph). So the DWG matrix model has two sets of couplings: original and dual.  The   partition function for graphs with spherical topology is then given by \eqref{QFTDWG} with all the weights \(w_{G}(\# v^*_{q_*},
\# v_{q})\equiv 1\). 

A particular case of this zero-dimensional DWG model was considered in the work~\cite{Kazakov:1996zm}.  For this case, the graphs are quadrangulations  and the vertices \(v_{2n}\)  can have arbitrary even orders \(2n\), i.e. with \(2n\)  neighboring square faces, where \(n=1,2,3\dots\), with the weights \(t_{2n}\)   chosen so that they suppress exponentially  the high orders \(n\gg 1\) of vertices. The partition function of such  quadrangulations  with spherical topology has been computed in ~\cite{Kazakov:1996zm}.

\begin{figure}[ht]
\begin{center}
\includegraphics[scale=0.35]{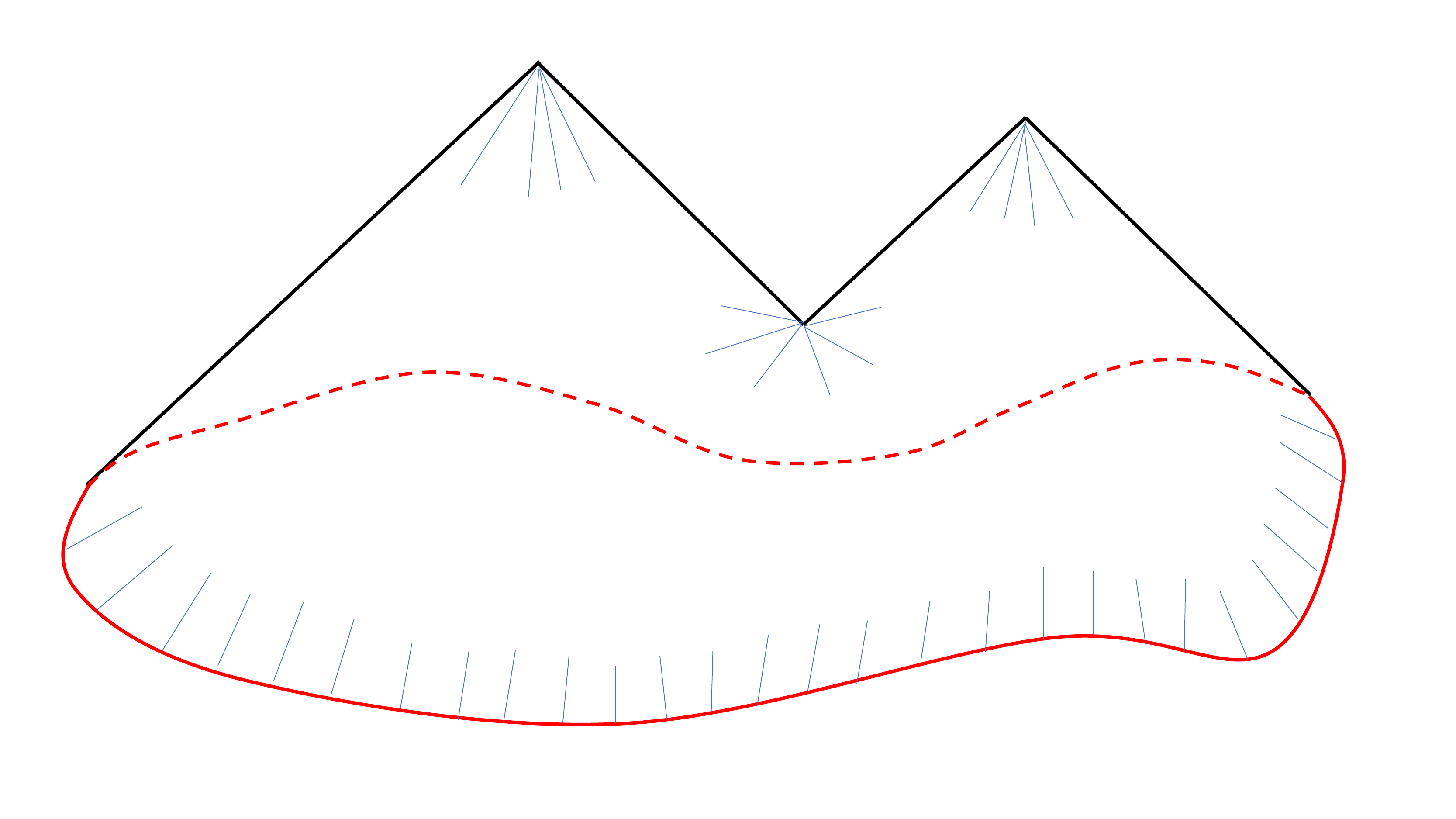}
\end{center}
\caption{ Schematic image of an abstract manifold with disc topology, with the boundary shown by the wiggled red line. It is flat everywhere except conical singularities. In this example we have two of such singularities with positive angle deficit (positive  curvature defects) and the 3rd, in the middle, with negative angle deficit (negative curvature defect). }
\label{fig:flat}
\end{figure}

In the current paper, we will study the same system of random quadrangulations but for the disc topology where we have  a single boundary of  even length \(L=2q\), as shown on figure~\ref{fig:flat}.  As will be explained below,  in order to make the problem analytically solvable, the boundary can be   introduced in  three different  ways. One particularly convenient way corresponds to cutting out from the closed quadrangulation  a vertex with \(L\) adjacent squares. The precise definition of that kind of disc partition function for this model is:
\begin{equation}\label{DWGdiscZ}
\mathcal{Z}(P,\lambda,\beta)=
        \sum_{L=2}^\infty L\,P^{-L}\sum_{Q_L}(\#v_L)\lambda^{F+1}\beta^{2(\# v_2-\delta_{L,2})}\,\ ,
\end{equation}    
where the first sum goes over the length \(L\) of the disc boundary (always even) and  the second sum goes over all disc quadrangulations \(Q_L\). Here \(\# v_k\) denotes the number of vertices of order \(k=2,4,6,\dots\) which correspond to  the angle deficits~\footnote{To define these angle deficits, we assume that all  faces are made of equal squares with the angles $\pi/2$ at each vertex.} \(\alpha_k=\pi (2-k/2)\). In the quantum Regge gravity picture~\cite{David:1984tx,Kazakov:1985ds,Kazakov:1985ea,Boulatov:1986jd,Kazakov:1988ch}, \(\alpha_{2}\)  corresponds to the single type of  positive  curvatures, \(\alpha_4\) introduces no curvature  and \(\alpha_{n>4}\) correspond to negative curvature insertions.   Furthermore, \(P\) is the length fugacity, \(\lambda\) is the area fugacity (the area \(F\) is defined as the number of square faces forming the quadrangulation), and \(\beta\) is the fugacity of positive curvature  (type \(\alpha_2\)) insertions.  Notice that due to the Euler theorem for the disc we have  \(\#v_2=\sum_{k\ge 3}\# v_k+{\rm const}\), so that the overall positive curvature is, up to an additive constant of the order $1$, equal to  the overall negative curvature, so that $\beta$  controls the overall scale (or the absolute value) of the total curvature. That allows us to interpolate between the regimes of large curvature fluctuations and the small curvature  regime dominated by almost flat (AF)  configurations which we describe in more detail later.

This model was solved exactly  in~\cite{Kazakov:1996zm} and it was called there the discretised \(2d\)~\(R^2\)~quantum gravity, since the possibility to control independently the positive and negative curvature amounts, in the continuous limit, to introducing irrelevant perturbations into the QG action, such as \(R^2\) (where \(R\) is the Gauss curvature), providing  a "flattening" effect on the geometry. In the continuum limit, it should lead  to the 2d QG action for the disc topology 
\begin{align}\label{R2action}
S_{R^2}^{\mathcal{M}}=-\frac{1}{2}\int_{\mathcal{M}} \sqrt{ g}\,\,(\tilde\lambda+\tilde\beta R^2)-\int_{\p\mathcal{ M}} \sqrt{h}\,(\tilde\gamma\,+\,K) \ .
\end{align}
Here we denoted by $\tilde\lambda,\tilde\beta,\tilde\gamma$ the bulk cosmological constant, the squared curvature coupling and the boundary cosmological constant, respectively  -- the renormalized continuous analogs of the lattice fugacities $\lambda,\beta,P$ in \eqref{DWGdiscZ}. We also used the notation $R$, $K$, $g$, and $h$ for the bulk and boundary gaussian curvatures  and metrics.

 The study of  the exact sphere partition function of this DWG model in the continuous limit (i.e. for large quadrangulations with no boundary) showed that there are there two smoothly connected regimes: almost flat manifolds (with very few conical singularities) and the pure $2d$  gravity regime of~\cite{David:1984tx,Kazakov:1985ds,Kazakov:1985ea,Boulatov:1986jd,Kazakov:1988ch}.  Since the $R^2$ coupling  $\tilde\beta$ has the dimension of length, the  former regime occurs for sufficiently small areas of manifolds, whereas the latter happens for large enough areas (the scale is set by $\tilde\beta$).

  In this paper, we study the quadrangulations having disc topology,  with the boundary of two different  types described in the next section.  We will keep the same continuous limit   of  quandrangulations with large area, but now also with  a  boundary of a fixed size (made out of a certain number of edges).  In the limit of large area, our  explicit results  interpolate between the universal  pure quantum gravity regime reproducing the well known disc partition function~\cite{Kazakov:1989bc}, and the regime of ``almost flat" (AF) random surfaces~\cite{Kazakov:1995gm}. The former regime assumes also the limit of a long boundary. In the  latter  regime, the surfaces are flat, consisting predominantly of    the vertices with four neighboring plaquettes that add no curvature,  everywhere except  rare insertions of conical singularities, with angle deficits multiple of $\pi$,   with specific exponential weights suppressing large angle deficits.   Thus in the continuous limit we simply sum up over the surfaces flat everywhere except a collection, or "gas"  of those conical defects. An intuitive image of such abstract flat manifold with curvature defects is presented on figure~\ref{fig:flat}.  
 
For the disc partition function of  quadrangulations with two different types of the boundary we found very similar, but not exactly identical expressions in the large area limit, given by eqs. \eqref{Wresy2} and \eqref{Wxi2}.  This similarity demonstrates certain universality features of this disc partition function, which however shows also a dependence on the boundary conditions. Both solutions interpolate  between the well known  pure 2d  QG regime and almost flat regime  for the disc partition function. In the former one, the results are completely universal, as expected.

 For the disc partition function in this nearly flat limit we found that there is no  phase transition between these two regimes, thus extending and supporting the findings of~\cite{Kazakov:1996zm} that were obtained at the level of the sphere partition function. While so far we did not observe signs of any more exotic phases in this DWG matrix model, such as the JT gravity regime~\cite{Saad:2019lba}, in the conclusions we  speculate on the possibility to find them for more sophisticated observables.  We will also discuss potential ways to produce from our model a discretized fluctuating geometry with the metric dominated by \(AdS_2\) background.

The paper is organized as follows.  In section \ref{sec:dwgch} we present the basic properties  of the general DWG model as well as defining the main observables we will study.  In subsection \ref{sec:defR2} there we define and discuss the particular $R^2$ QG matrix model studied in \cite{Kazakov:1996zm}. Then in section \ref{sec:characters} we return to the general DWG model and present the main steps towards its solution by character expansion, mainly reviewing the results of \cite{Kazakov:1995ae,Kazakov:1995gm}. From section \ref{sec:R2QG} onward we specialize to the $R^2$ QG model of \cite{Kazakov:1996zm}. In section \ref{sec:R2QG} we review its solution obtained in \cite{Kazakov:1996zm}. Then in section \ref{sec:critical} we present our main results. We compute a number of observables and explore their critical behavior. These are the resolvent for a particular kind of correlators with exponential accuracy, its continuum limit corresponding to nearly flat manifolds, and a similar limit for the resolvent describing the disc partition function  with a different boundary. We present conclusions in section \ref{sec:concl}, while the appendices contain a number of technical details.

\section{The DWG  matrix model via character expansion} 

\label{sec:dwgch}
 
The DWG matrix model is defined by~\cite{Das:1989fq,Kazakov:1995ae}
\begin{equation}\label{DWGMM}
 Z(t,t^*)=
\int {\cal D}M\exp \left[-\frac{N}{2}\tr  M^2+N\sum_{q=1}^{Q+1} \frac{1}{q}t_q\; \tr(MA)^q\right]\ ,
\end{equation}
where \(M\) is an \(N\times N\) Hermitian matrix integrated with the \(SU(N)\) invariant measure \(\mathcal{D}M\) and \(A\) is a fixed \(N\times N\) Hermitian matrix which can be always chosen diagonal without   loss of generality. 
The DWG matrix model~\eqref{DWGMM} cannot be solved directly by the diagonalization of the matrix \(M\). However, it turns out to be possible, in the large \(N\) limit, to solve it (in principle) by the character expansion methods worked out in~\cite{Kazakov:1995ae,Kazakov:1995gm,Kazakov:1996zm}\footnote{See also the recent work \cite{Mironov:2020tjf} the methods of which provide another way to the solution and perhaps to a generalization of the DWG models.}. In this section we will briefly review the main steps of this approach, referring the interested reader to those papers for the details. 

First we will  review the representation of $Z$ in terms of dually weighted graphs (DWG). Then we will  demonstrate the character expansion method which allows one to rewrite the model in terms of the sum over \(N\) highest weights of \(GL(N)\) representations (Young tableaux). At large \(N\)  the sum is dominated by a single large smooth Young tableau, and the computation of characters, as well as the saddle point equation, are reduced to a certain Riemann-Hilbert (RH) problem. The solution of this RH problem is directly related to three different types of disc partition functions of DWG, which are the main object of interest of this work.

\subsection{Dually weighted Feynman graphs} 

\begin{figure}[h]    
\includegraphics[angle=-90,scale=0.5]{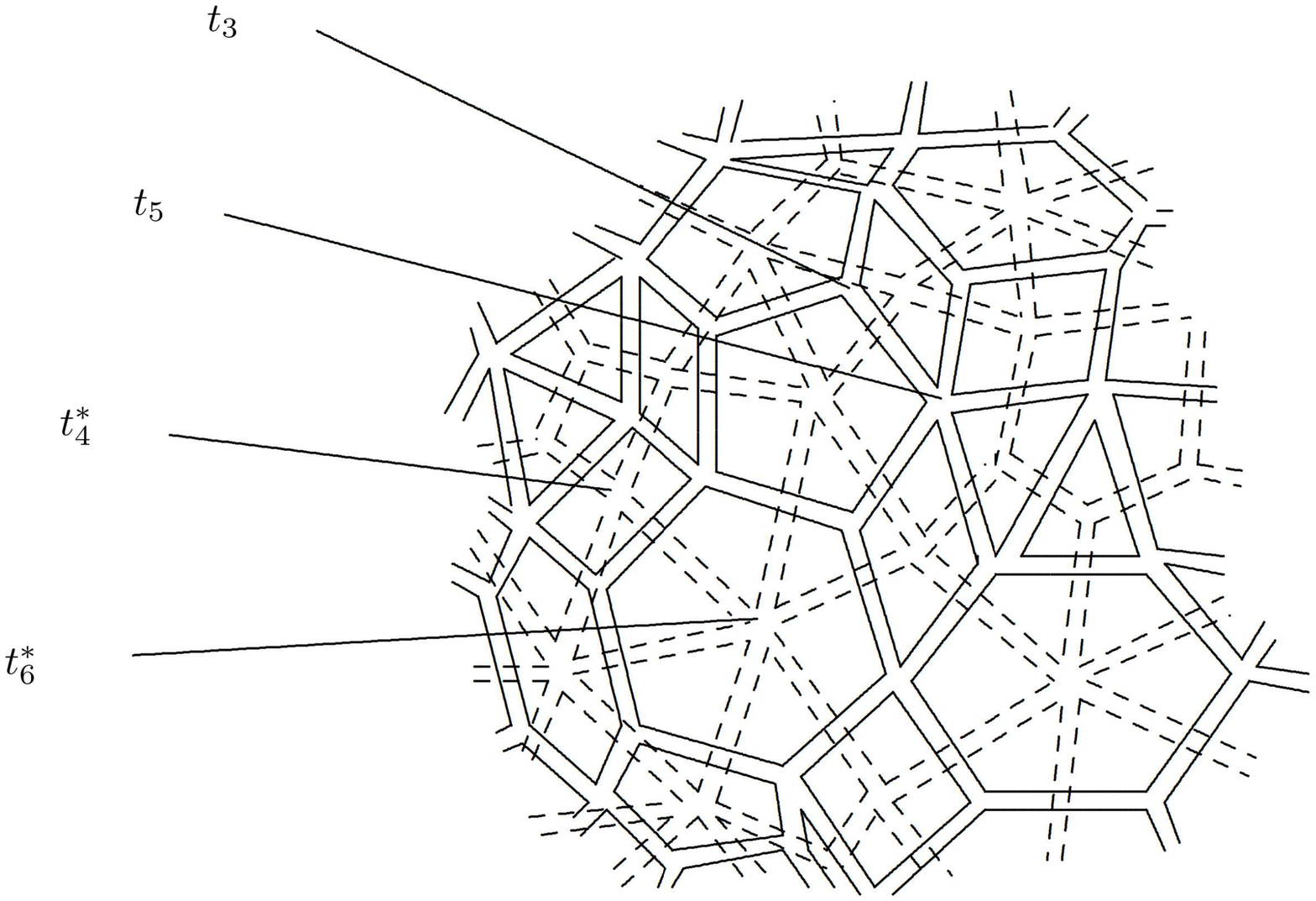}
\caption{Dually weighted graph (figure from \cite{Kazakov:1995ae}). We show the propagators as double lines. The couplings $t_n$ appear at $n$-vertices, while $t_n^*$ correspond to $n$-faces of the graph, corresponding to the faces of the dual graph depicted by dotted double-lines.}
\label{fig:DWGraph}
\end{figure}

The perturbative  expansion w.r.t. couplings \(t_q\)  is represented by planar  Feynman diagrams of the type  shown on  Fig.~\ref{fig:DWGraph}. Their elements are  the directed double-line propagators\footnote{"Directed" means here that two lines of the same propagator should be marked by opposite arrows, symbolising the covariant and contravariant indices. Since the direction in each index loop is conserved it ensures that each planar graph is orientable.} equal to 
\(\frac{1}{N}\delta_{i}^{j}\delta_{k}^{l}\) (the indices run from \(1\) to \(N\) and  are conserved along each line) and the vertices which give  \(Nt_q\,\,(A_{i_1j_1}A_{i_2j_2}\dots\, A_{i_qj_q})\). It is easy to see that at each face of order \(p\) of the lattice, after contraction of indices along each index line,  there will appear the factor
\beq
\label{tsdef}
    t_p^*=\frac{1}{N}\tr A^p\ .
\eeq
Hence the  partition function of this DWG matrix model has the following expansion in terms of DWG of a distinct topology~\cite{Kazakov:1995ae}:
\begin{equation}\label{ZDWGh}
\log Z(t,t^*)=
\sum_{G} N^{2-2g_G} \prod_{v^*_q,v_q \in G} {t_q^*}^{\# v^*_q}\ {t_q}^{ \#
v_q}\ ,\end{equation}
where the   sum goes over all Feynman graphs \(G, \) \(g_G\) is the genus of the graph \(G\), and the product goes over  all the vertices and faces of the graph. We indicate by $v_q$ vertices of order $q$ and by $v_q^*$ faces of order $q$, while $\# v_q$ and $\# v_q^*$ denote the number of these vertices and faces in a graph from the sum. Let us also note that we can view the $q$-faces as $q$-vertices on the dual graph $G^*$ (whose faces are the vertices of the original graph and vice versa).

From \eq{ZDWGh} we see that $Z$ is symmetric in the two sets of couplings $t_p\leftrightarrow t_p^*$, as there is a one-to-one correspondence between a graph $G$ and its dual $G^*$ . It is natural to introduce a `dual' matrix model with the same partition function but with Feynman diagrams corresponding to these dual graphs. It corresponds to interchanging all the couplings $t_p$ and $t_p^*$ in the original model and thus is defined by
\beq
\label{DWGd1}
    Z=\int {\cal D}\tilde M\exp \left[-\frac{N}{2}\tr  \tilde M^2+N\sum_{q=1}^{Q+1} \frac{1}{q}t^*_q\; \tr(\tilde MB)^q\right]\ ,
\eeq
where we have introduced a constant matrix $B$ which is the counterpart of $A$ satisfying a relation similar to \eq{tsdef},
\beq
\label{tpdef}
    t_p=\frac{1}{N}\tr B^p\ .
\eeq
We have also labelled the matrix being integrated over as $\tilde M$ instead of $M$ in the original model. Thus we see that in the dual model the roles of $A$ and $B$ are swapped. It is clear that such a duality, as well as the very representation  of (arbitrary) couplings in terms of traces of powers of these matrices can be in general possible only for \(N\to\infty\), at each order of topological \(1/N\) expansion. But later we will reformulate the model in terms of character expansion over $GL(N)$ representations (Young tableaux) where the characters will depend explicitly on independent couplings $t_n$ and $t_n^*$.

There exist three  matrix averages, computable by the developed methods, representing  the  DWG partition functions with a disc topology with three distinct  types of the boundary: 
\begin{align}
\cW_{L}&=\frac{1}{N}\langle\tr(AM)^L\rangle \ ,
\label{cWL}  
\\ 
\tilde \cW_{L}&=\frac{1}{N}\langle\tr(B\tilde M)^L\ ,\rangle
\label{cWLtilde}\\
W_{L}&=\frac{1}{N}\langle\tr M^L\rangle=\frac{1}{N}\langle\tr(\tilde M)^L\rangle\ ,  \label{bWL}
\end{align}
where the averages involving $\tilde M$ are computed in the dual model \eq{DWGd1}. The fact that the two types of correlators in \eqref{bWL} are equal may not be immediately obvious, but can be proven using the methods that we will discuss below.

The corresponding resolvents are 
\begin{align}
\cW(z)&\equiv\sum_{L\ge0}z^{-L-1}\cW_{L}=\langle\frac{1}{N} \tr \frac{1}{z-AM} \rangle \ ,
\label{cWR}  
\\ 
\tilde \cW(z)&\equiv\sum_{L\ge0}z^{-L-1}\tilde\cW_{L}=\langle\frac{1}{N} \tr \frac{1}{z-B\tilde M} \rangle\ ,&
\label{cWRtilde}\\
W(z)&\equiv\sum_{L\ge0}z^{-L-1}W_{L}=\langle\frac{1}{N} \tr \frac{1}{z- M} \rangle=\langle\frac{1}{N} \tr \frac{1}{z- \tilde M} \rangle \ . \label{bWR}
\end{align}

To realize the boundaries for \eqref{cWL} and \eqref{cWLtilde} as  the insertion of an extra \(L\)-vertex into the graph \(G\), we can simply differentiate the partition function w.r.t. the appropriate coupling:\footnote{There is no similar formula for \(W_L\) unless we generalize the matrix potential in \ref{DWGMM} by adding there \(\sum_{p\ge 3} \tau_p\, \tr M^p\).  Such a model, containing the 3rd infinite set of couplings,  is still solvable, in principle, by character expansion methods, and we would have \(W_L=\frac{1}{N^2}\p_{\tau_L}\log Z(t,t^*,\tau)\).}  \begin{align}\label{cWLlogZ}
\cW_{L}&=\frac{1}{N^2}\p_{t_L}\log Z(t,t^*)\ ,\\\label{cWLlogZtilde}
\tilde\cW_{L}&=\frac{1}{N^2}\p_{t^*_L}\log Z(t,t^*)\ .
\end{align}

  The corresponding disc partition functions have an obvious DWG picture. The first one~\eqref{cWLlogZ} can be viewed as a sphere partition function with one marked vertex of the order $L$  (divided by $L$, which takes away the overcounting due to the cyclic symmetry), for which the coupling $t_L$ is removed. The second one~\eqref{cWLlogZtilde} can be also viewed as a sphere partition function, but now with one marked face of the order $L$ (also divided by $L$) and its coupling $t_L^*$ removed. For the moments $W_L$ in the resolvent \eqref{bWR} the DWG interpretation is slightly more complicated: they can be also viewed as a sphere partition function with marked vertex of the order $L$ but since in the matrix model formulation there are no matrices $A$ around this vertex the adjacent faces do not contain it and are weighted each with the coupling $t_{n-1}$ for a face of order $n$. Obviously, all three resolvents \eqref{cWR}-\eqref{bWR} can be viewed as disc partition functions with such vertex or face removed.

Later we will explain that the quantities \eqref{cWR}-\eqref{bWR} can be computed directly from the saddle point equations for the DWG model.  

\subsection{Weighted quandrangulations (``\(R^{2}\) quantum gravity")}
\label{sec:defR2}

Before proceeding with key equations for the general DWG model, let us now introduce a particular case of the DWG model which we will focus on in this paper (starting from section \ref{sec:R2QG}). It was solved in~\cite{Kazakov:1996zm} and used there to study the \(R^2\) QG model. This model generates only  quadrangulations and has special  weights for the vertices, exponentially decaying with the order. This model will be of the main interest in this paper since  one can solve it explicitly for the disc partition functions and thus it is convenient for the search of various continuous limits of fluctuating 2d manifolds.  
 
The couplings of the model are given by
\beq
    t_n=\delta_{n,4} \ , \ \ \ \  t_n^*=T_n
\eeq
where $T_n$ are defined as
\begin{equation}
  \label{lambda-beta}
T_2=\frac{\lambda\beta^2}{\eps}\quad \text{and}\ \ \  T_{2q}=\lambda\eps^{q-2}\,\ \ , \ \ q\geq 2 \ .
\end{equation}
We see that they are all expressed in terms of 3 parameters: $\lambda,\beta$ and $\epsilon$. The partition function reads 
 \begin{equation}\label{DWGMMR3}
 Z=
\int {\cal D}M\exp N \tr \left(-\frac{1}{2}M^2
+\frac{1}{4}{ (AM)^{4}}\right)\ ,
\end{equation}
and for the dual matrix model we have
\begin{equation}\label{DWGMMR3d}
 Z=
\int {\cal D}\tilde M\exp N \tr \left(-\frac{1}{2}\tilde M^2
+\sum_{q=1}^{\infty} \frac{T_{2q}}{2q}{ (A_{4}\tilde M)^{2q}}\right)
\end{equation}
(as discussed above, these two partition functions are equal). 
Here the matrices $A$ and $A_4$ are chosen such that
\beq
\frac{1}{N}{\rm tr}A_4^n = \delta_{n,4} \ \ , \ \ \  \ \frac{1}{N}{{\rm tr}A^n}={T_n} \ .
\eeq
That also means that in our conventions we have
\beq
    B=A_4\ .
\eeq

Notice that the couplings $T_n$ were denoted by $t_n$ in \cite{Kazakov:1996zm}, which was a bit confusing since these are the couplings corresponding to faces, not vertices, in the original model \eq{DWGMMR3}. Thus we chose to introduce the new notation $T_n$ for them.

\subsubsection{Natural observables and combinatorics}

\label{sec:obs}

In our conventions it is the \textit{dual} model \eq{DWGMMR3d}, not the original model \eq{DWGMMR3}, that corresponds to graphs that are quadrangulations. Accordingly, we will study below the resolvent $\tilde\cW(z)$ generating the correlators $\langle\tr(A_4\tilde M)^n\rangle$ in this dual model, as well as the resolvent $W(z)$ that generates the $\langle\tr(\tilde M)^n\rangle=\langle\tr M^n\rangle$ correlators. Both observables correspond to graphs with disc topology built out of 4-vertex plaquettes. 

Invoking simple combinatorics and Euler theorem, the partition function in this dual model (equal, however, to the original $Z$) can be represented as \cite{Kazakov:1996zm}
\beq
   \log Z=\frac{\lambda^2\beta^8}{\epsilon^4}\sum_G \lambda^F \beta^{2(\#v_2-4)}\ ,
\eeq
where the sum is over connected graphs that are quadrangulations of the sphere.

We will be interested in particular in computing the resolvent $\tilde\cW$. The quantity \(\tilde\cW_{L}=\frac{1}{N}\langle\tr(A_4\tilde M)^L\rangle\) is given by graphs with the shape of a  quadrangulated disc with the boundary of  (even) length \(L\) corresponding to a (removed) vertex of order \(L\). So, to get \(\tilde{\cal\ W}_L\), the weight of the vertex given by  \(T_L=\lambda\eps^{L/2-2}\)  should be removed
from the partition function 
\eqref{ZDWGh}. Explicitly, we see from \eq{DWGMMR3d} that
\beq
    \frac{1}{N}\langle {\rm Tr}(\tilde MA_4)^L\rangle = L\frac{\d \log Z}{\d T_L}\ ,
\eeq
and \eq{ZDWGh} implies that taking the derivative in $T_L$ in the sum over graphs simply produces a factor $\propto \# v_L$. As in our case from \eq{lambda-beta} we have $T_L=\lambda\epsilon^{L/2-2}\beta^{2\delta_{L,2}}$ (for even $L$), from the equations above we get
\beq
    \frac{1}{N}\langle {\rm Tr}(\tilde MA_4)^L\rangle=\sum_{G_L}L \#v_L\lambda^{F+1}\epsilon^{-2-L/2}\beta^{2\#v_2-2\delta_{L,2}}\ ,
\eeq
where the sum is over connected graphs with at least one $L-$vertex.
This gives for the resolvent
\beq   
\label{Wec}
    z\tilde{\cal W}(z)=\sum_{L\geq 0}z^{-L}\frac{1}{N}\langle {\rm Tr}(\tilde MA_4)^L\rangle=1+
        \frac{1}{\epsilon^2}\sum_{L\geq 2,\;L\ {\rm even}} \left(\epsilon^{1/2}z\right)^{-L}f_{L}(\lambda,\beta)\ ,
\eeq 
where 
\begin{align}
\quad f_{L}(\lambda, \beta)= \lambda L\beta^{-2\delta_{L,2}}\sum_{G_L}(\#v_L)\lambda^F\beta^{2\# v_2}\,\ .
\end{align}
We see that the dependence of the result on $\epsilon$ is very restricted, namely up to a $1/\epsilon^2$ factor it is a function of $\epsilon^{1/2}z$.
 
Via an argument of the same type one can obtain a similar combinatorial formula for the correlators $\langle{\rm Tr} (MA)^L\rangle$. At the same time, for the $\langle {\rm Tr}M^L\rangle$ correlator the combinatorics seems to be more complicated and we leave its exploration for the future.

Below in section \ref{sec:characters} we continue describing the main results regarding the general DWG models. Then from section \ref{sec:R2QG} we will specialize to the case of the $R^2$ QG model \eq{DWGMMR3} we have just discussed here.

\section{Character expansion and planar limit for DWG model }
\label{sec:characters}

In this section, we will review  the reduction of the DWG matrix model integral to the character expansion over Young tableaux. Then we will use the fact that the sum over Young tableaux goes over only $N$ highest weights and apply the saddle point approximation.  

\subsection{DWG partition function as a sum over $SU(N)$ representations}

The way to represent the partition function of the DWG matrix model  is as follows~\cite{Kazakov:1995ae}:  first we expand the exponent of the second term in the action~\eqref{DWGMM} w.r.t. the Schur characters \(~\chi_{R}[t^*]\)  and \(GL(N)\) characters \(\chi_{R}[AM]\), where \(R\) is a representation of \(GL(N)\), then  we use the orthogonality property of the characters to integrate over the angular variable $\Omega\in SU(N)$ in the angular decomposition $M=\Omega^\dagger X\Omega$, where $X={\rm diag}\{x_1,x_2,\dots,x_N\}$ is the diagonal matrix of eigenvalues, and finally we perform the explicit Gaussian integral over \(M\). The details are reviewed in the appendix~\ref{CharExp_sec}. In this way we obtain the following formula for the DWG partition function in terms of the sums over representations:\footnote{Strictly speaking the r.h.s. of \eq{Zh} includes an extra factor $N^{N(N-1)/4-1/2\sum_k h_k}$ (see \cite{Kazakov:1995ae}) but it can be reabsorbed into a redefinition of the matrices $A,B$ by a scalar factor}  
\begin{equation}\label{Zh}
Z(t,t^*)=c\,\sum_{\{h^e,h^o\}}
\frac{\prod_i(h^e_i-1)!!h^o_i!!}{
\prod_{i,j}(h^e_i-h^o_j)}~\chi_{\{h\}}[t]~\chi_{\{h\}}[t^*]\ .
\end{equation}
Here the sum goes over  \(GL(N)\) representations labeled by Young tableaux  defined by the shifted highest weights $h_j$,
\begin{equation} 
\label{hjdef}
Y=\{h_j=N-j+m_j,\quad j=1,2,\dots,N\}\ ,
\end{equation} 
where $m_j$ is the number of boxes in $j$'th row, with equal numbers of even and odd shifted highest weights \(\{h_1,\dots, h_{N/2},h_{N/2+1},\dots,h_{N}\}\equiv\{h_1^e,\dots, h_{N/2}^e,h_1^o,\dots,h_{N/2}^o\}\)~\footnote{The highest weights are not necessarily ordered here according to the values, but they can always be reordered so due to the antisymmetry of the characters}. The formula \eq{Zh} was obtained by pure combinatorics of planar graphs in~\cite{DiFrancesco:1992cn} and rederived in~\cite{Kazakov:1995ae} from the DWG matrix model.
 The factors \(\chi_{\{h\}}[t],\ \chi_{\{h\}}[t^*]\) here are usual Schur characters -- polynomials of couplings \(t_k,t_k^*\). One way to define them is through Schur polynomials $P_n(\theta)$ which are read off from
\beq
    e^{\sum_{n=1}^\infty z^n\theta_n}=\sum_{n=0}^\infty z^nP_n(\theta)\ ,
\eeq
where
\beq
\label{thed}
    \theta_k=\frac{1}{k}{{\rm tr}\; B^k}=\frac{N}{k}t_k\ ,
\eeq
and for $n<0$ we define $P_n(\theta)=0$. Then we have
\beq
\label{chiP}
    \chi_{\{h\}}(t)=\det_{k,l}(P_{h_k+1-l}(\theta))\ ,
\eeq
and similarly for $\chi_{\{h\}}(t^*)$ where we use $A$ instead of $B$ in \eq{thed}.

Notice that the representation  \eqref{Zh}  of the DWG matrix model partition function renders the symmetry between two sets of couplings \(t_q\leftrightarrow t_q^*\) obvious.

A clear advantage of the representation  \eqref{Zh}  is the drastic reduction of the number of variables: instead of \(N^2\) original matrix variables \(M_{ij}\) it contains only the sums over \(N\)  highest weights \(h_j\). Hence, to sum up DWG's we can apply the saddle point approximation to find the dominating large Young tableau for the sums in~\eqref{Zh}. But for that we have to learn how to compute the characters in this limit.  The corresponding methods, as well as the formulas of the following subsection, have been worked out in~\cite{Kazakov:1995ae,Kazakov:1995gm,Kazakov:1996zm,Kazakov:1996vy}.

\subsection{Characters in the large $N$ limit}

Using the methods of the papers mentioned above, we can compute the Schur characters depending on arbitrary number of variables (couplings)   at large \(N\) (for large smooth Young tableaux with characteristic shifted highest weights $h_j\sim N$).   The computation is based on  simple identities for Schur characters. The first one is 
\begin{equation}\label{tchi=chi}
t_{L}\,\cdot\chi_{\{h\}}[t]=\frac{1}{N}\sum_{j=1}^{N}\chi_{\{h+L\delta_j\}}[t]\,,\qquad  \end{equation}  
where in the r.h.s. one adds \(L\) to one of the highest weights $h_j$ in the Young tableau and sums up the resulting character over all \(N\) insertions\footnote{The highest weights can then be reordered inside the character, using the permutational symmetry, to satisfy the inequalities {\(h_{j}> h_{j+1} \)} following from the definition of $h_j$ in \eq{hjdef}.} so that
\beq
    \{h+L\delta_{j}\}\equiv \{h_i +L \delta_{i,j},\quad i=1,2,\dots, N\}\ .
\eeq
The second one is 
\begin{align}\label{dchi}
 \frac{L}{N}\p_{t_L}\chi_{\{h\}}[t]~=~\sum_{j=1}^{N} \,{\chi_{\{h-L\delta_{j}\}}[t]}\ ,
\end{align} 
where in the r.h.s. one subtracts \(L\) from one of the highest weights in the Young tableau. While in the r.h.s. here we may potentially get negative values of the weights $h_k-L$ defining the character, this identity still holds\footnote{For a different definition of the character the identity may not hold when $L$ is large enough. However, in what follows we will focus on the large $N$ limit and the saddle point distribution of $h_k$ for which $h_k\sim N$, so for fixed $L$ the combination $h_k-L$ is almost always positive and does not cause any problems.} with the definition \eq{chiP} of the characters where we take $P_n=0$ for $n<0$.

Using \eqref{tchi=chi} 
  we also have, denoting the Vandermonde determinant by $\Delta$   (see the eqs.(2.2)-(2.3) of~\cite{Kazakov:1995ae}),  
\begin{align}
t_L&= \frac{1}{N}\sum_{k=1}^{N}\frac{\chi_{\{h+L\delta_{k}\}}[t]\,}{\chi_{\{h\}}[t]}
=\frac{1}{N}\sum_{k=1}^{N}\frac{\Delta(h+L\delta_{k})}{\Delta(h)}e^{\log\frac{\chi_{\{h+L\delta_{k}\}}}{\Delta(h+L\delta_{k})}-\log\frac{\chi_{\{h\}}}{\Delta(h)}}=  \notag\\
  &=\frac{1}{N}\sum_{k=1}^N\prod_{j\ne k}\left(1+\frac{L}{h_k-h_j}\right)e^{L{\mathbf F}(h_k)}+{\cal O}(1/N)=\frac{1}{L}\oint \frac{dh}{2\pi i}e^{L(H(h)+{\mathbf F}(h))}+{\cal O}(1/N)
\label{tHF}\end{align}
   where  we  exponentiated the product in the large \(N\)  limit (see \cite{Kazakov:1995ae} for details), when the characteristic  \(h_j\sim N \),  introducing two basic functions:
   the resolvent of the highest weights
 \begin{equation}\label{resH}
 H(h)=\sum_{j=1}^{N}\frac{1}{Nh-h_j}\equiv\int \frac{dh'\,\rho(h')}{h-h'}\ .
 \end{equation}
 and
\begin{equation}\label{Ffunction}
{\mathbf F}
(h)=\p_{h_k}\left.\log\frac{\chi_{\{h\}}[t]}{\Delta(h)}\right|_{h_k=Nh}\ .
\end{equation}
We also introduced here the normalized highest weight variable \(h\sim {\cal\ O}(N^0)\). One can also define a function similar to $\bF$ but for the dual weights $\{t_n^*\}$,
\beq
\label{Fsdef}
    \bF^*(h)=\p_{h_k}\left.\log\frac{\chi_{\{h\}}[t^*]}{\Delta(h)}\right|_{h_k=Nh}
\eeq
and we get a relation similar to \eq{tHF},
    \begin{align}
    \label{tsHF}
t_L^*&=\frac{1}{L}\oint \frac{dh}{2\pi i}e^{L(H(h)+{\mathbf F}^*(h))}\ .
\end{align}

Let us point out that our notation here differs somewhat from \cite{Kazakov:1995ae}, as there our function $\bF^*$  was denoted by $F$. Furthermore the notation for $F$ differs between \cite{Kazakov:1995ae} and subsequent papers \cite{Kazakov:1995gm,Kazakov:1996zm}. We chose to use bold font here to avoid confusion.  We will use $\bF,\bF^*$ here in section \ref{sec:characters} while we discuss the general solution, and then we will use $F$ in later sections when we specialize to the $R^2$ DWG model discussed in \cite{Kazakov:1996zm} (see section \ref{sec:even} for a summary of the notation in the latter case).

Using \eqref{dchi} and the first line of \eqref{chichi} given in appendix \ref{CharExp_sec} we  also derive a similar formula \cite{Kazakov:1995ae,Kazakov:1995gm} for \(\langle\tr(B\tilde M)^L\rangle\) evaluated in the dual model, 
\begin{align}\label{tildeMB}
 \frac{1}{N }\langle\tr( B\tilde M )^L\rangle &=\frac{1}{L}\oint_{C_H} \frac{dh}{2\pi i}e^{-L[H(h)+\bF^*(h)]}\,
\end{align} 
where for large $h_j\sim N$ we also used the fact that  the sum over representations reduces at the saddle point to a single, dominating Young tableau defined by the resolvent $H(h)$. For the correlator $\langle\tr(MA)^{L}\rangle$ we find in a similar way the representation
\begin{align}\label{Amcorr1}
 \frac{1}{N }\langle\tr( A M )^L\rangle &=\frac{1}{L}\oint_{C_H} \frac{dh}{2\pi i}e^{-L[H(h)+\bF(h)]}\, \ . 
\end{align} 
There is also an alternative formula for the same quantity, given in \cite{Kazakov:1995ae},
\beq
\label{MAcorr}
    \langle\frac{1}{N}\tr(AM)^{2L}\rangle=\frac{1}{L}\oint\frac{dh}{2\pi i}h^{L}e^{L(H(h)+2\bF^*(h))}
\eeq
which however we will not use in this paper.

In order to compute  \(\langle \tr M^L \rangle\) one can also use similar arguments (see appendix \ref{CharExp_sec} for more details). As a result we get \cite{Kazakov:1995ae} 
\beq
\label{Mcorr}
   \langle\frac{1}{N}\tr M^{2L}\rangle=
   \frac{1}{L}\oint \frac{dh}{2\pi i}h^Le^{LH(h)}\ .
\eeq

\subsection{Saddle point equation and general RH problem for DWG }

From now on, we will consider the DWG models only with polynomial original and dual potentials, with the highest powers equal to \(Q\) and \(Q^*\), respectively, so that the potentials read
\beq
    V(M)=\sum_{q=1}^{Q} \frac{t_q}{q}M^q \ , \ \ \ V_*(M)=\sum_{q=1}^{Q^*} \frac{t^*_q}{q}M^q\ .
\eeq
The large  \(N\) saddle point approximation for the multiple sum \eqref{Zh} then reads
\begin{equation}\label{spZh}
\,
~\p_h\left[\frac{1}{2}h_k\log h_k-\frac{1}{2}\log\Delta(h)+\log\chi_{\{h\}}[t]~+\log\chi_{\{h\}}[t^*]\right]=0\,.
\end{equation}
We used here natural assumptions~\cite{vershik1981asymptotic,Douglas:1993iia,Kazakov:1995ae} that the densities of distributions of even and odd highest weights \(\{h^{e}_j\)\} and \(\{h^{o}_j\}\) are the same: \(\rho^{e}(h)=\rho^{o}(h)=\frac{1}{2}\rho(h)\). We also applied the Stierling formula \(\log h!\simeq h\log h \) valid for typical \(h_j\sim N \) at the saddle point.
Using the definitions \eqref{Ffunction} and \eqref{resH} we can rewrite \eqref{spZh} in the continuous form. For the DWG model with general sets of original and dual couplings we get:
\begin{align}
2\bF+2\bF^{*}+3\sH+\ln h=0 \label{DWGspe}
\end{align}
where by slash we denote the symmetric part of the function \(H\) on its cut.

To be precise, here we need to discuss (following \cite{Kazakov:1995ae}) an important property of the density $\rho(h)$. The density is naturally defined  on the interval $h\in [0,a]$ going from zero to some  endpoint which we denote as $a>0$. In fact $\rho(h)$ is saturated at its maximal\footnote{the constraint $\rho\leq 1$ follows from the definition of the density and the fact that the shifted highest weights $h_i$ are strictly decreasing with $i$ as follows from \eq{hjdef}} value $\rho(h)=1$ at the start of this interval, i.e. for $h\in [0,b]$ with $b<a$. Accordingly, $H(h)$ has a logarithmic cut for $h\in[0,b]$. As discussed in \cite{Kazakov:1995ae} the saddle point equation \eq{DWGspe} holds only on the remaining part of the interval where $H$ has a nontrivial cut, i.e. for $h\in[b,a]$ (with \(H(h\pm i0)=\sH(h)\mp i\pi\rho(h)\)).

The saddle point equation should be  supplemented by  equations for the character functions  \(\bF\) and \(\bF^*\)  defining the characters as functionals of this highest weight distribution. Such equations were derived in~\cite{Kazakov:1995ae} from the observations described in the previous subsection.
Namely, we introduce the functions \begin{equation} \bG(h)=e^{\bF(h)+H(h)}\,,\qquad \bG^*(h)=e^{\bF^*(h)+H(h)}\ .\,\end{equation}
Notice that like for $\bF,\bF^*$ our notation differs from the one of \cite{Kazakov:1995gm,Kazakov:1996zm} as there $G$ denoted a quantity closely related to our $\bG^*$ (see section \ref{sec:even} for more details on the notation).

The functions \(\bG(h)  \) and \(\bG^{*}(h)\) have \(Q\) and \(Q^{*}\) sheets, respectively. As the eqs.\eqref{tHF}, \eqref{tildeMB} and \eqref{Amcorr1}  show, the behavior at zero  for each one is given by expansions
\begin{align}
h(\bG)&=\sum_{q=1}^{Q}t_{q}\bG^{-q}+\sum_{q=0}^{\infty}\bG^{q}\langle\frac{1}{N}\tr(A M)^q\rangle=\bG^{-1}V'(\bG^{-1})+\bG^{-1}\, \cW(\bG^{-1})\label{hGexp}\\
h(\bG^*)&=\sum_{q=1}^{Q^{*}}t^{*}_{q}\bG^{*-q}+\sum_{q=0}^{\infty}\bG^{*q}\langle\frac{1}{N}\tr(B\tilde M)^q\rangle=\bG^{*-1}V_{*}'(\bG^{*-1})+\bG^{*-1}\,\tilde\cW(\bG^{*-1})
\label{hG*exp}
\end{align} 
where \(\tilde M\) is the matrix for the dual matrix model and we put \(t_0=t_0^*=1\) while $V'(M)$ denotes the derivative of the potential. We assumed here that all the small $\bG$ singularities come from the original matrix potentials.  

These expansions show that the functions \(\bG(h)\) and \(\bG^*(h)\) have  branch points of order \(Q \) and \(Q^*\) at \(h\to \infty\).  This suggests another form of  eqs. \eqref{hb1} and \eqref{hG*exp}  (see eq. (3.23) of \cite{Kazakov:1996zm}):
\begin{align}\label{CharRH}
\frac{(-1)^{Q-1}}{t_{Q}}\prod_{q=1}^{Q}\bG_{q}=\frac{1}{h}e^{H(h)}=\,\frac{(-1)^{Q^{*}-1}}{t_{Q^{*}}}\prod_{q=1}^{Q^{*}}\bG^{*}_{q}
\end{align}  where by \(\bG_q\) we denoted the values of the function $\bG$ 
on \(Q\) different sheets, and similarly for \(\bG^*.\)
If we skip the label we always mean the main sheet \(\bG=\bG_{1}\).

The equations \eqref{DWGspe}, \eqref{CharRH}, together with the small $\bG,\bG^*$ asymptotics \eqref{hb1}, \eqref{hG*exp}, give the complete Riemann-Hilbert problem for the DWG model with general sets of \(Q\) original and \(Q^* \) dual couplings.

\section{The \(R^2\) quantum gravity solution}
\label{sec:R2QG}

While above we discussed general DWG models, from now on we will specialize to the particular case of DWG model described in section~\ref{sec:defR2} and studied in \cite{Kazakov:1996zm}. In this section  we will review its exact solution from \cite{Kazakov:1996zm}.  The model corresponds to quadrangulations with special, exponential weights for the vertices.  We keep for it the name ``\(R^2\)~QG" model given in~\cite{Kazakov:1996zm}. The main advantage of this model, apart from the fact that the solution for $H(h)$ can be given explicitly in terms of elliptic functions, is that the couplings are chosen in such a way that we have a parameter controlling the local "curvature" fluctuations (number of nearest neighbors for the vertices of quadrangulations) together with another one  controlling the area (number of squares in the quadrangulation). Thus we can nontrivially restrict the geometry of the graphs and find interesting continuum limits.

In the paper \cite{Kazakov:1996zm} the sphere partition function of such quadrangulations has been studied. In the current paper we will generalize this analysis to the \textit{disc} partition functions. Namely, we will compute in the near-critical regime the corresponding resolvents $\tilde\cW(z)$ and $W(z)$ defined above. In this section we summarise the exact solution which serves as the starting point for our new computation.

\subsection{Notation and key relations for the case of even potentials}
\label{sec:even}

From now on we will follow the notation of \cite{Kazakov:1996zm} which is tailored for the case (of which the $R^2$ QG model is an example) where the potentials for both the original and the dual models are even, so that $t_{2n+1}=t^*_{2n+1}=0$. In this case we have $\tr A^{2n+1}=0$ and one can parameterize the matrix $A$ as
\beq
    A=\begin{pmatrix}
        \sqrt{a} & 0
        \\ 0 & -\sqrt{a}
    \end{pmatrix}
\eeq
where $a$ is an $N/2\times N/2$ matrix. Then instead of \eq{Fsdef} one uses the function $F$ given by
\beq
\label{Fe}
    F(h_k)=2\frac{\d}{\d h_k^e}\log\frac{\chi_{\{\frac{h_e}{2}\}}(a)}{\Delta(h_e)}
\eeq
so it is defined in terms of only the even weights $h_k^e$ with $k=1,\dots,N/2$~\footnote{We assume that, by symmetry, both even and odd weights are distributed in the same way}, and the character of the matrix $a$.
It can be deduced from \cite{Kazakov:1995gm,Kazakov:1996zm} that it is related to our notation as\footnote{The difference between $\bF^*$ and $F$ comes mainly from the fact that in \eq{Fe} we have the Vandermonde of only the even weights $h_k^e$, while in \eq{Fsdef} we have the Vandermonde for all the weights. }
\beq
F(h)=2\bF^*(2h)+2H(2h)-H(h)\ .
\eeq
Moreover the function $G$ used in \cite{Kazakov:1995gm,Kazakov:1996zm} differs from our $\bG,\bG^*$ and is defined as
\beq
    G(h)=e^{H(h)+F(h)}\ .
\eeq
Then analogs of equations \eq{tsHF}, \eq{tildeMB} read\footnote{Strictly speaking one should rederive these equations for the even case as the contour of integration used in the trick in \eq{tHF} becomes subtle to choose due to overlapping cuts (see \cite{Kazakov:1995gm} for the derivation and comments on this). }
\beq
    t_{2q}^*=\oint \frac{dG}{2\pi i G}h(G)G^q\ ,
\eeq
and 
\beq
    \frac{1}{N}\langle \tr (B\tilde M)^{2q}\rangle=\oint \frac{dG}{2\pi i G}h(G)G^{-q}\ ,
\eeq
so we have
\beqa
\label{heven}
h(G)&&=1+\sum_{q=1}^{Q}t_{2q}G^{-q}+\sum_{q=1}^{\infty}G^{q}\langle\frac{1}{N}\tr(B \tilde M)^{2q}\rangle
\label{hGexp2}
\\ \nn
&=&1+\sum_{q=1}^{Q}t_{2q}G^{-q}+\frac{1}{\sqrt{G}} \tilde\cW\left(\frac{1}{\sqrt{G}}\right)\,.
\eeqa
Note that for the $R^2$ QG model we have $B=A_4$ where $A_4$ is the matrix discussed in section \ref{sec:defR2} with the property $\frac{1}{N}\tr A_4^n=\delta_{n,4}$.

Let us also note that in the case of even potential we find from \eq{Mcorr} that \cite{Kazakov:1995ae}
\begin{align}\label{resWH}
zW(z)\equiv\langle\ \frac{1}{N} \tr \frac{z}{z-M} \rangle=1-\oint_{H} \frac{dh}{2\pi i}\log \left( z^2-h^{}e^{\,H(h)}\right)\ .\end{align} 
By the Lagrange inversion formula (we have a single cut on the main sheet of \(H(h))\) we get from here the parameterisation of the resolvent as\footnote{For   \(\cW(z)\), which in the even case is given by (see \cite{Kazakov:1995ae})
\begin{align}
z\cW(z)\equiv\langle\ \frac{1}{N} \tr \frac{z}{z-AM} \rangle=1-\oint_{C_H} \frac{dh}{2\pi i}\log \left( z^2-he^{2\bF^*(h)+H(h)}\right) \ ,\end{align}
the same inversion trick unfortunately does not work  since  the integration contour \(C_H\) here is pinched between the cuts of \(H(h)\) and  \(\bF^*(h)\).}  \begin{align}
\label{Wpar1}
zW(z)=z^{2}-h\,,\qquad  z^{2}=h^{}e^{\,H(h)}
\end{align}

\subsection{Explicit solution for the density}

To derive the saddle point equation for this case, the easiest way is to plug into \eqref{Zh} the character $\chi_{\{h\}}[t]$ in the form~\eqref{Schur4}. This gives for the partition function
\begin{equation}\label{Zh4}
Z(t,t^*)=\,\sum_{\{h^e,h^o\}}
\frac{\prod_i(h^e_i-1)!!h^o_i!!}{
\prod_{i,j}(h^e_i-h^o_j)}~\prod_{\epsilon=0}^{3}\frac{\Delta(h^{(\epsilon)})}{
\prod_i\bigl(\frac{h^{(\epsilon)}_i-\epsilon}{ m}\bigr)!}
\,\,{\rm sgn}\bigl[\prod_{0\leq\epsilon_1<\epsilon_2\leq3}
\prod_{i,j}(h^{(\epsilon_2)}_i-h^{(\epsilon_1)}_j)\bigr]~\chi_{\{h\}}[t^*].
\end{equation}
Here following \cite{Kazakov:1996zm} we introduced four groups with equal numbers  of weights $h^{(\epsilon)}_i,\quad \eps=0,1,2,3,\,\,i=1,\dots,N/4$ such that $h^{(\epsilon)}_i=\epsilon \mod 4$. We will assume that all four groups  are distributed with the same density and the ${\rm sgn}[\dots]$ factor is irrelevant in the large $N$ limit. That provides the saddle point equation given in eq.(4.7) of \cite{Kazakov:1996zm}. It reads
\begin{equation}
\label{saddlepointeq}
2F+\sH+\log h=0\,,\,\qquad  h\in[b,a]. 
\end{equation}
where by $\sH(h)$ we defined the principal value of the resolvent $H(h)$ on the cut $C_H=[b,a]$, \(H=\sH\mp i\pi\rho(h)\).

To complete the RH problem defining two functions -- the resolvent $H(h)$ and the character function  $F(h)$ -- first we notice that with our definition of the couplings $t_n^*$ of the dual potential \eq{lambda-beta}, we obtain from \eqref{hG*exp} 
\begin{align}\label{htog}
h-\frac{\lambda\beta^2}{\epsilon }\frac{1}{G}-\frac{\lambda G}{(G-\eps)} = \frac{1}{\sqrt{G}}\tilde\cW(\frac{1}{\sqrt{G}})
\end{align}
where we used the definition of the resolvent:  $\tilde\cW(z)\equiv\sum_{q =0}^{\infty}\langle\tr(\tilde MA_4)^{2q}\rangle z^{-2q-1} $.  We notice that for $G\to 0 $ the r.h.s.  of the last formula is vanishing since $\tilde\cW(z)\simeq 1/z$ for $z\to \infty$. The vanishing l.h.s. then suggests that the function $G$ has only two sheets connected through the cut of  $F(h)$, so that we can put $Q=2$ in \eqref{CharRH}. Then the logarithm of equation \eqref{CharRH} looks as follows:
\begin{align}
\log G(h+ i0)+\log G(h- i0)-H(h)+\log \frac{h}{(\beta^2-1)\lambda}=0\ ,\qquad h\in[d,c]\,.   
\label{GpmH}\end{align}
which gives, on the defining sheet of $F(h)$,
the second needed RH equation 
\begin{align}
2\sF(h)+H(h)+\log \frac{h}{(\beta^2-1)\lambda}\,=0\qquad h\in[d,c]\,.   
\label{RR2}\end{align}
We used the fact that $F(h)$ has a cut $C_F:\,[d,c]$ on its defining sheet, whereas   $H(h)$ does not have a $[d,c]$ cut.

Remarkably, both equations \eqref{saddlepointeq} and \eqref{RR2}  of this RH problem turn out to be written for the same function $2F(h)+H(h)+\log h$ which thus has only two sheets. This helps to immediately write its solution in terms of the Cauchy integrals (see eq.(4.8) in~\cite{Kazakov:1996zm}) which can be expressed in terms of incomplete elliptic integrals. Since by virtue of \eqref{saddlepointeq} we also have 
\begin{align}
    2F(h)+H(h)+\log h=\,i\pi \rho
\end{align}
it immediately gives us the explicit formula for the density of the highest weights of the saddle point Young tableau.

In order to write the solution explicitly, let us introduce some notation. 
 We define the elliptic moduli $k^2$ and $k'^2$ corresponding to the branch points as
 \beq
 \label{kdef}
    k^2={\frac{(a-b)(c-d)}{ (a-c)(b-d)}},\qquad k'^2=1-k^2 \ , \ 
 \eeq
 and we denote by $K$ and $K'$ the complete elliptic integrals of the first kind with moduli $k$ and $k'$ respectively. The corresponding elliptic nome $q$ and its dual $q'$ read 
  \begin{equation}\label{nom}
   q=e^{-\pi K'/K} \ , \ q'=e^{-\pi K/K'}\ .
\end{equation}
We also describe the Mathematica conventions for some of these functions (and others that we use below) in appendix \ref{app:ell}. In this notation the result for the density found in \cite{Kazakov:1996zm} reads
\begin{equation}
\label{density}
\rho(h)=
\frac{u}{ K} - \frac{i}{ \pi}~\ln \Bigg[
\frac{\theta_4\big(\frac{{\pi}}{2 K} ( u - i v),q \big)}
 {\theta_4\big(\frac{\pi}{2 K} ( u + i v),q \big)} \Bigg]
\end{equation}
where $u$ is defined in terms of the inverse Jacobi elliptic function sn,
\beq
\label{defu}
    u={\rm sn}^{-1}\left(\sqrt{\frac{(a-h)(b-d)}{(a-b)(h-d)}},k\right)\ .
\eeq

In the next subsection we describe equations that fix the branch points $a,b,c,d$ in terms of the couplings of the model.

\subsection{Fixing parameters of the solution}
\label{sec:parameters}

The branch cut endpoints $a,b,c,d$ are fixed in terms of the couplings $\lambda,\beta$ and $\epsilon$ by a system of rather nontrivial equations found in \cite{Kazakov:1996zm}. It is obtained by imposing constraints such as correct normalization of the resolvent at large $h$. We refer the reader to \cite{Kazakov:1996zm} for details of the derivation, and here we present the final result, which we also rewrote in a slightly more compact form.

Let us first introduce some additional notation, namely
\beq
\label{vsdef}
    v=\sn^{-1}\left(\sqrt{{\frac{a-c}{ a-d}}},k'\right) \ , \ \ s=\sn( v,k')\ .
\eeq
In addition, we define
\beq
    \Delta\equiv b-c\ .
\eeq
Then the equations that fix $a,b,c,d$ read \cite{Kazakov:1996zm}
\begin{align}
v&=-K'-\frac{K}{\pi}\log (\lambda(1-\beta^2))
\label{vtolkb}\\
\label{lameq}
\lambda&=\frac{ 2 K q }{  i\pi^2}~
\frac{\theta_1\big(\frac{i \pi v }{ K} ,q\big)}{ \theta'_1\big(0,q\big)}
\bigg(-E+K\left(k'^2s^2+\frac{\Upsilon\Xi }{s}+2\,\Xi^2\right)\bigg)
\\
\label{Deltaeq}
\Delta&= \frac{ 4 K q}{ \epsilon^2i\pi }~
\frac{\theta_1\big(\frac{i \pi v }{ K} ,q\big)
}{ \theta^{'}_1\big(0,q\big)}\,\frac{k'^2 s\sqrt{   
    1-s^2}}{\sqrt{1 -k'^2 s^2}},
\\
\label{bm1eq}
b-1&=\frac{1}{\epsilon^2}\left[ \frac{  2Kq }{ i\pi}~
\frac{\theta_1\big(\frac{i \pi v }{ K} ,q\big)
}{ \theta^{'}_1\big(0,q\big)}\left(
\frac{k'^2s^4-2s^2+1}{s\sqrt{1-s^2}\sqrt{1-k'^2s^2}}+~
2\Xi\right)-2\lambda\right]
\end{align}
where
\beq
\label{Updef}
\Upsilon=
\frac{3k'^2s^4-2(k'^2+1)s^2+1}{\sqrt{1-s^2}\sqrt{1-k'^2s^2}} \ , \ \ \Xi=\frac{\pi} {2K}+E(v,k')+\left({\frac{E}{K}}-1\right)~v
\eeq
and $E$ is the complete elliptic integral of the second kind (similarly to $K$) while $E(v,k')$ denotes the incomplete elliptic integral of the 2nd kind (see appendix \ref{app:ell} for the corresponding Mathematica notation).

Let us elaborate briefly on the structure of these four equations \eq{vtolkb}-\eq{bm1eq}. Our goal is to find $a,b,c,d$ as functions of the three couplings $\lambda,\beta,\epsilon$. We see that plugging $v$ from the first equation into the second one gives an equation that completely fixes $k^2$ in terms of two couplings $\lambda$ and $\beta$. The first equation then fixes $v$, and from the last two equations we find $b$ and $c$. Finally, having found $k$ and $v$ and recalling their definitions \eq{kdef}, \eq{vsdef} we fix the remaining two parameters $a,d$. 

Using  identities between elliptic functions, one can rewrite these equations in a variety of ways. Here we presented the form we found the most useful for our calculation.

\subsection{Simplification and computation of  \(G\)} 
\label{sec:solution}

Below we will be interested in computing the resolvent $\tilde\cW$ which is encoded in $G(h)$ and thus it is useful to write the latter function explicitly. Since  \(2F+H+\log h=\pm i\pi\rho\) we first get (choosing the sign accordingly)
\begin{align*}
 &2\log G(h)+\log h=(2F+H+\log h)+H= -i\pi\rho+\int_0^a  \frac{dx}{h-x}\rho(x)\ .
 \end{align*}  
 We found it useful to also write this result in a slightly different way. Notice that we can get rid of the term without theta functions in the density \eq{density} by switching from $\theta_4$ to $\theta_1$ using the identity
\beq
\label{theta41}
    \theta_4\big({\pi \over 2 K}  u +\frac{i\pi}{2}\frac{K'}{K},q \big)=
    \frac{i}{q^{1/4}}\exp\left(-\frac{i\pi u}{2K}\right)\theta_1\big({\pi \over 2 K}  u ,q \big)\ .
\eeq
This gives
\begin{equation}
\label{density1}
\rho(h)=-{i \over \pi}~\ln \Bigg[-
{\theta_1\big({\frac{\pi}{2 K}} ( u - i (v+K')),q \big)
\over
 \theta_1\big({\frac{\pi}{2 K}} ( u + i (v+K')),q \big)} \Bigg]\ .
\end{equation}
In addition, the form of the density suggests one to rewrite the integral defining $H(h)$ in terms of $u$ (defined in \eq{defu}) rather than $h$. Let us denote by $w(h')$ the same function as in \eq{defu} but with $h$ replaced by $h'$. Then in the definition of $H(h)$ given by \eq{resH} we change the integration variable $h'\to w(h')$ and obtain
\begin{align}
\label{Hresu}
H(h)&=-\frac{2i \,k^2\left(1 -k'^{2}s^2\right)}{\pi k'^{2} }
\int_{0}^{K}\frac{   dw\,\,\cn(w) \dn(w) \sn(w)\,\log \left[-
\frac{\theta_1\big({\pi \over 2 K} ( w - i (v+K')),q \big)
}{\theta_1\big({\pi \over 2 K} ( w + i (v+K')),q \big)}\right] }{(s^2 \dn(w)^2-1) ^2\left[\frac{h-b}{\Delta}+\frac{1-\frac{1}{k^{'2}} \dn(w)^2}{1-s^2 \dn(w)^2}\right]}\ .
\end{align}
We can also write the argument $h$ as a function of $u$ by inverting \eq{defu},
\begin{align}
    \label{hu2}
\frac{h-b}{\Delta}= \frac{ 1 }{(k^2-1) s^2}-\frac{ (k^2-1) s^2+1}{(k^2-1) s^2 (k^2 s^2\, \sn(u)^{2}-s^2+1)}\ .
\end{align}
Then finally for $G(h)$ we have
\beqa
\label{Gint1}
    2\log G(h)&=& \nn
    -\frac{2i \,k^2\left(1 -k'^{2}s^2\right)}{\pi k'^{2} }
\int_{0}^{K}\frac{   dw\,\,\cn(w) \dn(w) \sn(w)\,\log \left[-
\frac{\theta_1\big({\pi \over 2 K} ( w - i (v+K')),q \big)
}{\theta_1\big({\pi \over 2 K} ( w + i (v+K')),q \big)}\right] }{(s^2 \dn(w)^2-1) ^2\left[\frac{h-b}{\Delta}+\frac{1-\frac{1}{k^{'2}} \dn(w)^2}{1-s^2 \dn(w)^2}\right]}
 \\ \label{Gint}
    &-&i\pi\rho(h)-\log(h-b)\ .
\eeqa
This is the form of $G(h)$ that we will use below.

\subsection{Scaling properties of $G$ as a function of $\epsilon$}

\label{sec:Geps}

Let us show that dependence of $G(h)$ on $\epsilon$ is quite trivial. From \eq{heven} we see that to obtain the resolvent \(\tilde{\cal W}(z)=\langle\tr\frac{1}{z- \tilde MA_4}\rangle\) we need to compute the function \(G(h)\) and invert it. Denoting
\beq
    z\equiv \sqrt{\epsilon/G}
\eeq 
we have
\beq\label{htW}
    h=z^2\frac{\lambda\beta^2}{\epsilon^2}+\frac{\lambda z^4}{\epsilon^2(1-z^2)}+\frac{z}{\sqrt{\epsilon}}\tilde{\cal W}[(\sqrt{\epsilon}/z)^{-1}]\ .
\eeq
On the other hand, as we saw above from the combinatorial argument leading to \eq{Wec}, the resolvent should have the form
\beq
\label{Wi1}
    \tilde{\cal W}(z)=\frac{1}{z}\left(1+\epsilon^{-2}\phi_1(\sqrt{\epsilon}z)\right)
\eeq
where the function $\phi_1$ does not depend on $\epsilon$ and $z$. Plugging this into \eq{htW} we get
\beq
    \epsilon^2(h-1)={\lambda\beta^2z^2}+\frac{\lambda z^4}{1-z^2}+\phi_1(z)\ .
\eeq
Inverting this relation we find
\beq
    G=\epsilon\phi_2(\epsilon^2(h-1))
\eeq
where $\phi_2$ is again some function which does not depend on $\epsilon$ or $h$ explicitly. As a result, this predicts for $G$ the scaling property
\beq
\label{Gesc}
    G(h,Z\epsilon )=ZG(Z^2(h-1)+1,\epsilon)\ .
\eeq

Let us verify that the result for $G$ given above in \eq{Gint1} indeed satisfies this relation. We see that in the equations \eq{vtolkb}-\eq{bm1eq} which define parameters of the solution the variable $\epsilon$ appears only in $\Delta$ and in one term in the last equation \eq{bm1eq}. Using this, one can easily check that all terms in \eq{Gint1} except the last one are functions of the combination $\epsilon^2(h-1)$ and thus sending $\epsilon\to Z\epsilon$ for them is the same as replacing $h\to Z^2(h-1)+1$. The very last term $\log(h-b)$ in \eq{Gint1}, on the other hand,  produces an extra $\log Z$ term under this transformation. As a result we see that \eq{Gesc} is perfectly satisfied! 

Due to this scaling symmetry we see that $\epsilon$ cannot affect any critical behavior properties of the resolvent $\tilde{\cal W}$. Accordingly, for the purpose of studying them we will set $\epsilon=1$ in the subsequent computations in sections \ref{sec:exp} - \ref{sec:limits}. Then in section \ref{sec:otheres} we will study a different kind of resolvent, namely $W(z)$, for which the dependence on $\epsilon$ is more tricky. There we will restore the $\epsilon$-dependence in the needed equations, which is not hard to do as we will see.

  \section{Critical regimes of large area and boundary of the quadrangulated disc }
\label{sec:critical}

In this section, we will discuss the critical regimes for the partition function of the quadrangulated disc. Namely, we will first tune the parameter $\lambda$  in the resolvents $\tilde\cW(z)$ and $W(z)$ for the elliptic solution \eqref{density}  of sections~\ref{sec:parameters}, \ref{sec:solution}  to its critical value $\lambda_c$ in such a way that the area (number of plaquettes in the graph)  becomes very large. This allows one to consider the typical quadrangulations which "forget" about the details of their discretization and can be considered as relatively smooth 2d manifolds with fluctuating metric. Generically, this regime leads to the well known solution of pure 2d quantum gravity~\cite{David:1984tx,Kazakov:1985ds,Kazakov:1985ea}, as it was demonstrated in~\cite{Kazakov:1996zm}. However, to study the transition between this pure 2d QG regime and the ``almost flat" regime, when the curvature is maximally suppressed, we have an extra parameter to adjust, namely $\beta$ which controls the fluctuations of the modulus of overall curvature accumulated on the disc. Tuning $\beta\to0$ one can "flatten" the disc in such a way that the vertices with four neighbors will dominate, with rare insertions of the conical curvature defects, having angle deficits $+\pi$ (positive curvature defect) or $-\pi,-2\pi,-3\pi,\dots$ (negative curvature defects). By appropriate simultaneous tuning of $\lambda\to\lambda_c$ and $\beta\to 0,$ with a certain double scaling, this limit interpolates between the regime of pure 2d gravity, when the the size of the manifold is large enough to accommodate the "pure gravity"  fluctuations of the metric,    and "almost flat" (AF) regime when the size of the manifold is relatively small and thus it is almost completely flattened, apart from  rare positive conical curvature defects (all negative curvature is then concentrated at the boundary). 
\footnote{The quantisation unit of the angle deficit $\delta\phi =\pi$ should be an important relevant  parameter  in such double scaling limit, and the result would be different for a different value of $\delta\phi$.  It would be interesting to solve a model with different $\delta\phi$, say for triangulations.
}  

The universal partition function  interpolating between two such regimes  was established in~\cite{Kazakov:1996zm}.  We will extend in this section the analysis of  this  ``$R^2$ gravity" solution of the DWG model to the disc partition function. Namely, we will explicitly  calculated the resolvents $\tilde{\cal W}(z)$ and $W(z)$  in this  ``near-flat"  regime -- large area of quadrangulations but arbitrary length of the boundary -- which is the main result of this paper. Then we will study  these quantities in  the two  limits mentioned above. In the limit of pure gravity, which demands another double scaling computation, adjusting the boundary cosmological constant \(z\to z_c\) simultaneously with \(\lambda\to\lambda_c\), we will reproduce for both resolvents $\tilde{\cal W}(z)$ and $W(z)$  the well known universal disc amplitude~\cite{,Kazakov:1989bc}. In the limit of almost flat 2d manifolds with minimal number of curvature defects  we will reproduce the  disc partition function from~\cite{Kazakov:1995gm} (see eq.(4.17) there).    

As a preliminary step, we will first investigate a special regime for the general  solution,  when the elliptic nome $q$ goes to $1$ but we drop only the exponentially small terms. It is convenient to parametrize $q$ as
\begin{equation}
    q=e^{-\pi\tau}
\end{equation}
so that $\tau\to 0$ and we work with exponential accuracy, i.e. up to ${\cal O}(e^{-\pi/\tau})$ terms. Let us recall that all parameters of the solution~\eqref{density} for the density of highest weights $\rho(h)$ are fixed in terms of two original parameters of the model $\lambda$ and $\beta$. In practice it will be convenient for us  to first deal with intermediary parameters  $v$ (defined in \eqref{vsdef}) and $\tau$  instead. We will  compute $G(h)$ at generic $v$ and with exponential precision in $\tau$. We present this calculation in subsection \ref{sec:exp}. 

Next, in subsection \ref{sec:crit2} we recall (following \cite{Kazakov:1996zm}) the main features of the critical regime in our model, and in particular the case when $v$ is small (in addition to $q\to 1$). We will refer to this regime as the `flattening' limit. In  subsection \ref{sec:resflat} we will  describe the properties of our solution in this regime in detail, making contact with the results in \cite{Kazakov:1996zm} and presenting a number of nontrivial checks of the result. We also discuss how the solution interpolates between the pure gravity and the AF regimes. Finally in subsection \ref{sec:otheres} we compute $W(z)$, i.e. the resolvent for $\langle {\rm Tr}M^n \rangle$ expectation values,  study its properties and reproduce the same pure gravity limiting behavior for it as well.

\subsection{Solution with exponential precision for near-flat regime}
\label{sec:exp}

We will focus here on the  limit $\tau\to 0$
which was explored for the partition function of the model in \cite{Kazakov:1996zm}.
 Here we will compute the function $G(h)$ in this limit with exponential precision in $\tau$. We will keep the second remaining parameter $v$ finite so that we do not yet specialize to the vicinity of the critical line discussed in \cite{Kazakov:1996zm} (we will do that in the next subsection).

\subsubsection{Expanding the parameters}
\label{sec:exppar}

Let us first compute how the cut endpoints $a,b,c,d$ and related parameters of the solution behave in the limit $\tau\to 0$. With exponential precision in $\tau$ we have
\begin{equation}
    K\simeq \frac{\pi}{2\tau} \ , \  E(k)\simeq 1
\end{equation}
while
\begin{equation}
    {\rm sn}(v,k')\simeq\sin v \ , \ E(v,k')\simeq v\ .
\end{equation}
Plugging this into \eqref{Updef} we find
\begin{equation}
    \Upsilon\simeq\frac{\cos 2v}{\cos v} \ , \ \Xi\simeq \tau+\frac{2\tau v}{\pi}\ .
\end{equation}
We also need to expand the theta functions appearing in \eqref{Deltaeq}. Using a modular transformation to change their modulus from $q$ to $q'$, we find with exponential precision 
\begin{equation}
    \frac{\theta_1(\frac{i\pi v}{K},q)}{\theta_1'(0,q)}\simeq i \tau  e^{\frac{4 \tau  v^2}{\pi }} \sin (2 v)\ .
\end{equation}
Then \eqref{Deltaeq} gives the result for $\Delta$,
\begin{equation}
\label{Dexp}
    \Delta\equiv b-c\simeq16e^{-\pi/\tau}e^{\tau(4v^2/\pi-\pi)}\sin^22v
\end{equation}
and we get from \eqref{lameq}, \eqref{bm1eq} 
\begin{equation}
\label{lbexpe}
    \log(\lambda(1-\beta^2))=-\tau(2v+\pi)
\end{equation}
\begin{equation}
\label{lamqq}
    \lambda\simeq \frac{e^{\frac{4 \tau  v^2}{\pi }-\pi  \tau } \left(\left(\tau  (2 v+\pi )^2-\pi \right) \sin (2 v)+\pi  (2 v+\pi ) \cos (2
   v)\right)}{\pi ^2}
\end{equation}
\begin{equation}
\label{brese}
    b\simeq 1+\frac{e^{\frac{4 \tau  v^2}{\pi }-\pi  \tau } \left(2 \left(-4 \tau  v^2-2 \pi  \tau  v+\pi \right) \sin (2 v)-4 \pi  v \cos (2
   v)\right)}{\pi ^2}
\end{equation}
\begin{equation}
a\simeq 1+
\frac{e^{\frac{4 \tau  v^2}{\pi }-\pi  \tau } \left(4 \sin (v) \left((\pi -2 \tau  v (2 v+\pi )) \cos (v)+\pi ^2 \sin (v)\right)-4 \pi 
   v \cos (2 v)\right)}{\pi ^2}
\end{equation}
\begin{equation}
\label{dexpe}
    d\simeq 1+\frac{e^{\frac{4 \tau  v^2}{\pi }-\pi  \tau } \left(2 (\pi -2 \tau  v (2 v+\pi )) \sin (2 v)-4 \pi  \left(\pi  \cos ^2(v)+v \cos (2
   v)\right)\right)}{\pi ^2}\ .
\end{equation}
The first two of these relations are equivalent to those given in equation (5.5) of \cite{Kazakov:1996zm} \footnote{Notice the second relation in (5.5) from \cite{Kazakov:1996zm} contains a typo, it should read $q=(\lambda(1-\beta^2))^{1/(1+2v/\pi)}$.}. 
 We also find that the relation between $h$ and $u$ from \eqref{defu} in our regime becomes
\begin{equation}
\label{hue}
    h-b=\frac{e^{\tau(4v^2/\pi-\pi)}\sin^22v}{\cosh^2 u-\sin^2 v}\ .
\end{equation}

\subsubsection{Computing the integral for $G(h)$}
\label{sec:integrals}

The nontrivial part of $G(h)$ we need to compute is the integral in \eqref{Gint1}. The term with theta functions in the integrand is the density \eqref{density1} which we can write with exponential precision as
\begin{equation}
\label{rhoe}
    \rho=-\frac{i}{\pi}\log\left[-\frac{\theta_1\big(\frac{\pi}{ 2 K} ( u- i (v+K')),q \big)
}{\theta_1\big(\frac{\pi}{ 2 K} ( u+ i (v+K'),q \big)}\right] \simeq
\frac{4\tau}{\pi^2  }u(v+\pi/2)-\frac{i}{\pi}\log\left(\frac{\cosh\big( u -iv \big)
}{\cosh\big(u+iv\big)}\right)\ .
\end{equation}
To write the rest of the integrand in our limit we can expand it directly, or, more conveniently, first recall that the integral we need is simply  $\int_b^a\frac{dx}{h-x}\rho(x)$ which we can rewrite with $w$ as the integration variable,
\begin{equation}
\label{intba}
    \int_b^a\frac{dx}{h-x}\rho(x)=-\int_0^Kdw\frac{dh(w)}{dw}\frac{1}{h(u)-h(w)}\rho(w)\ .
\end{equation}
Then we can compute $dh(w)/dw$ with exponential accuracy in our limit $\tau\to 0$ from \eqref{hue}, which gives
\begin{eqnarray}
\label{dhe}
    \frac{dh(w)}{dw}\frac{1}{h(u)-h(w)} &=& -\frac{\sinh w}{\cosh w+ \cosh u}-\frac{\sinh w}{\cosh w- \cosh u}
    \\ \nn
    &+&  \frac{\sinh w}{\cosh w+ \sin v} 
    +\frac{\sinh w}{\cosh w- \sin v} \ .
\end{eqnarray}
Notice also that in our integral \eqref{intba} we can (with exponential precision) replace  the upper integration limit $K$ by $\infty$ and use the approximation for the density \eqref{rhoe} on the whole region of integration, as we justify in detail in appendix \ref{app:exp}.

From the expression for the density \eqref{rhoe} we see that there are two types of integrals we need to compute, corresponding to the two terms in that equation.  The first type of integrals reads\footnote{The proper choice of branch we use here corresponds to ${\rm Im}\; u<0$ and real $v>0$.}
\beq
   \int_{0}^\infty dw\frac{wh'(w)}{h(u)-h(w)}=u^2+i \pi  u+v^2-\frac{\pi ^2}{4}
\end{equation}
which is relatively straightforward to compute in Mathematica. The second type of integrals is
\begin{equation}
   I=- \int_0^\infty dw\frac{h'(w)}{h(u)-h(w)}\log\frac{\cosh(w+iv)}{\cosh(w-iv)}\,.
\end{equation}
 It is much more nontrivial but can still be computed with the help of a number of tricks. The initial result we found in Mathematica contains many ${\rm Li}_2$ polylogarithms, which can then be further reduced using specialised software (see e.g. \cite{Maitre:2005uu,Panzer:2014caa})\footnote{We thank \"Omer G\"urdo\(\check{\rm g}\)an for help with simplifying the result for this integral.}.  The final outcome reads
\begin{eqnarray}
    && 
    I=-\text{Li}_2\left(-e^{2 i v-2 u}\right)-\text{Li}_2\left(-e^{2 u+2 i v}\right)-2 \text{Li}_2\left(1-e^{2 i v}\right)-2
   \text{Li}_2\left(1+e^{2 i v}\right)
   \\ \nn &&
   -2 u^2+4 i u v+2 v^2+\frac{\pi ^2}{3}
   \\ \nn &&
   + (-2 u-2i v-i\pi ) \log \left(1+e^{2 u+2 i v}\right)+(2 u-2 i v+i \pi ) \log \left(e^{2
   u}+e^{2 i v}\right)
   +2 i \pi  \log \left(1+e^{2 i v}\right)\ .
\end{eqnarray}

Combining all the parts together and using \eq{lbexpe}, \eq{lamqq}, we get the final result for $G$,
\begin{equation}
\label{Grese}
    G=(\lambda(1-\beta^2))^{-\frac{1}{2}+\frac{v}{\pi}}\frac{\sqrt{\cosh^2u-\sin^2v}}{\sin 2v}{\rm exp}(\Phi)
\end{equation}
where
\begin{eqnarray}\label{Phi} && \nn
    \Phi=u^2 \left(-\frac{\tau  (2 v+\pi )}{\pi ^2}-\frac{i}{\pi }\right)+u \left(-\frac{4 v}{\pi }-\frac{2 i \tau  (2 v+\pi )}{\pi
   }\right)+\frac{\tau  (\pi -2 v) (2 v+\pi )^2}{4 \pi ^2}
   \\  &&
   +\frac{1}{6} i \left(\frac{18 v^2}{\pi }-12 v+\pi \right)-\log \left(1+e^{2 i v}\right)
   \\ \nn &&
   +\frac{(i u+v-\pi ) \log \left(1+e^{2
   u-2 i v}\right)}{\pi }+\frac{(-i u+v+\pi ) \log \left(1+e^{2 u+2 i v}\right)}{\pi }
   \\ \nn &&
   -\frac{i}{2\pi}
   \left[
   \text{Li}_2\left(-e^{2 i v-2 u}\right)
   +\text{Li}_2\left(-e^{2 u+2 i v}\right)
   +2\text{Li}_2\left(1-e^{2 i v}\right)
   +2\text{Li}_2\left(1+e^{2 i v}\right)
   \right] 
   \end{eqnarray}
In view of \eq{lbexpe}, \eq{lamqq} we can exclude $\tau$ so this last equation can be also written as
\begin{eqnarray}\label{Phitt} && \nn
    \Phi=u^2 \left(
    \frac{\log(\lambda(1-\beta^2))}{\pi ^2}-\frac{i}{\pi }\right)+u \left(-\frac{4 v}{\pi }
    +\frac{2 i \log(\lambda(1-\beta^2))}{\pi
   }\right)
   \\  &&
   -\frac{\log(\lambda(1-\beta^2)) (\pi^2 -4 v^2)}{4 \pi ^2}
   +\frac{1}{6} i \left(\frac{18 v^2}{\pi }-12 v+\pi \right)-\log \left(1+e^{2 i v}\right)
   \\ \nn &&
   +\frac{(i u+v-\pi ) \log \left(1+e^{2
   u-2 i v}\right)}{\pi }+\frac{(-i u+v+\pi ) \log \left(1+e^{2 u+2 i v}\right)}{\pi }
   \\ \nn &&
   -\frac{i}{2\pi}
   \left[
   \text{Li}_2\left(-e^{2 i v-2 u}\right)
   +\text{Li}_2\left(-e^{2 u+2 i v}\right)
   +2\text{Li}_2\left(1-e^{2 i v}\right)
   +2\text{Li}_2\left(1+e^{2 i v}\right)
   \right] 
   \end{eqnarray}
We remind that $h$ and $u$ are related in our limit via \eqref{hue}, into which can also substitute \eq{lbexpe} to get
\begin{equation}
\label{hue2}
     h-b=(\lambda(1-\beta^2))^{1-\frac{2v}{\pi}}\frac{\sin^22v}{\cosh^2 u-\sin^2 v}\ .
\end{equation}
We also recall that $v$ can be expressed in terms of $\lambda,\beta$ from \eq{lbexpe}, \eq{lamqq} which give the equation fixing it, 
\beq
\lambda=\frac{\left(\lambda(1 -\beta ^2)  \right)^{1-\frac{2 v}{\pi }} \left(\pi  (2 v+\pi ) \cos (2 v)-\sin (2 v) \left((2 v+\pi ) \log
   \left(\lambda(1-\beta ^2) \right)+\pi \right)\right)}{\pi ^2}
\eeq

Thus we have two equations \eqref{Grese} and \eqref{hue2} which give $G$ and $h$ as functions of $u$ and therefore implicitly determine $G(h)$. It would be interesting to explore  various critical regimes possibly hidden in this rather nontrivial function. We will recover in the next subsection the so called `flattening' regime,  leaving a more in-depth exploration  for future work.

\subsection{Critical regime and flattening limit}

\label{sec:crit2}

The critical regime dominated by large quadrangulations for this matrix model was studied in \cite{Kazakov:1996zm}. It was understood that it corresponds to the case when the derivative of $\d\rho/\d h$ vanishes at the endpoint $a$. This gives the following additional constraint on the parameters:
\begin{equation}
\label{Xicrit}
   1=\frac{{\rm cn}(u,k')}{{\rm sn}(u,k'){\rm dn}(u,k')}\Xi\ .
\end{equation}
This equation allows one to express $\lambda$ (and all other parameters such as $v$, $b$ etc) as a function of $\beta$. As a result we have a 1-parametric critical line. Let us also mention that in the $q\to 1$ regime with exponential precision considered above in section \ref{sec:exp}, the condition \eqref{Xicrit} for criticality further simplifies and reduces to (see \cite{Kazakov:1996zm})
\begin{equation}
    \tan v=-\frac{\log(\lambda(1-\beta^2))}{\pi}\ .
\end{equation}

A particularly important part of the parameter space is the region when $\lambda\to 1$ and $\beta\to 0$. This also implies that we are close to the critical line (which passes through the point $\lambda=1,\beta=0$) though not necessarily directly on it. This is the  limit discussed in \cite{Kazakov:1996zm} which we call `flattening'. As described there it corresponds to $\tau, v$ and $\beta$ all being small and of the same order $\sim\beta$. We find from \eq{lbexpe}, \eq{lamqq} that up to $\beta^2$ corrections we have in this limit
\begin{equation}
    q\simeq \lambda
\end{equation}
and 
\beq
\label{vxd}
    v=\frac{\beta}{\sqrt{2}}(x-\sqrt{x^2-1})\ .
\eeq
Here we introduced (following \cite{Kazakov:1996zm}) the variable
\beq
\label{defxl}
    x=1+\frac{\sqrt{2}(\lambda_c-\lambda)}{\pi\beta}
\eeq
which is kept finite in our `flattening' limit, and we also define
\beq
\label{deflc}
    \lambda_c=1-\frac{\pi\beta}{\sqrt{2}}
\eeq
which corresponds to the critical value of $\lambda$ (to first order in $\beta$). It is also convenient to write \eq{vxd} as
\beq
\label{vV}
    v=V\beta \ , \ \ \ \text{with} \ \ V=\frac{1}{\sqrt{2}}(x-\sqrt{x^2-1})
\eeq
so that $V$ is a finite parameter which becomes equal to $1/\sqrt{2}$ on the critical line. We will write the results in terms of $V$ and $\beta$ in what follows. From the equations above we also find 
\begin{equation}
\label{tauv}
    \tau\simeq\beta\frac{2V^2+1}{4V}\ .
\end{equation}

Let us remind that $\Lambda=2(\lambda_c-\lambda)$  plays the role of bulk cosmological constant controlling the area (number of plaquettes) where as $\beta$ controls the curvature fluctuations.

 In the next subsection we will discuss the properties of the solution for $G(h)$ and the resolvent $\tilde\cW$ in this regime.

\subsection{Resolvent for ${\rm Tr}(A_4\tilde M)^n$ correlators in the flattening limit}

\label{sec:resflat}

The flattening limit affects the solution significantly as the cut structure becomes partially degenerate. From the results given in section~\ref{sec:exppar} we find that the cut endpoints $a,b,c,d$ approach the values
\begin{equation}
    a,b,c\simeq 1 \ , \ \ \  d\simeq -3
\end{equation}
so that the $[a,b]$ cut collapses to a point and moreover almost touches the $[c,d]$ cut. The distance between the two cuts (i.e between $b$ and $c$) is exponentially small as we see from \eqref{Dexp} that $b-c\sim e^{-\pi/\beta}$, while the size of the $[a,b]$ cut is of order $\sim \beta^2$.

 Expanding our result \eqref{Grese} in this regime we get
\begin{equation}
\label{Gb1}
   G=-\frac{\sin r}{2V\beta}\left(1+\frac{\beta  r (r-\pi )}{4 \pi  V}-\frac{\beta  V \left(-r^2+\pi  r+2 (\pi -2 r) \cot (r)+4\right)}{2 \pi }+O(\beta^2)\right)
\end{equation}
and
\begin{equation}
\label{hb1}
    h=b+\frac{4\beta^2V^2}{\sin^2 r}(1-\beta\pi\frac{2V^2+1}{4V} )+O(\beta^4)\ .
\end{equation}
Here to make the result more compact we introduced the variable $r$ related to $u$ as
\begin{equation}
    u=-ir-i\pi/2\ .
\end{equation}

Our goal is to express $h$ in terms of $G$ and extract the resolvent $\tilde\cW$ from \eqref{heven}. From the structure of \eqref{Gb1} we see that when $\tau\to 0$ we can keep finite the combination 
\begin{equation}
    y=2\beta VG
\end{equation}
as well as $r$. Notice that in order to write the result for $\tilde\cW(P)$ we should take in \eqref{heven} $G=1/P^2$ which gives
\beq
    y=2V\beta/P^2 
\eeq
so that we keep $\beta/P^2$ fixed while $\beta,P\to 0$. Then writing
\begin{equation}
    r=r_0+\beta r_1+\dots
\end{equation}
and expanding \eqref{Gb1} at small $\beta$ we find $r(y)$ perturbatively, e.g.  $r_0=-\arcsin y$. Plugging the result then into \eqref{hb1}, and using that in our regime \eqref{brese} gives
\begin{equation}
    b=1+\frac{2 \beta ^3 V \left(2 V^2-3\right)}{3 \pi }+O(\beta^4)
\end{equation}
we find to order $\beta^3$
\begin{eqnarray} \nn
    h-1&=&\frac{4\beta^2V^2}{y^2}+\frac{2 \beta ^3 V \left(2 V^2-3\right)}{3 \pi }
    +
    \beta ^3 V \frac{ \left(2 V^2+1\right)\left(2 \sin ^{-1}(y) \left(\sin ^{-1}(y)+\pi \right)- \pi ^2\right)}{\pi  y^2 }
    \\
    \label{hsery} &+&
    \beta^3\frac{8 V^3 \left( \sqrt{1-y^2}(
   \pi +2  \sin ^{-1}(y))-2 y\right)}{\pi  y^3 }\ .
\end{eqnarray}
Now we can write explicitly the resolvent itself, given by 
\begin{equation}
    \sum_{n=1}^\infty G^n\langle\frac{1}{N}\tr(\tilde MB)^{2n}\rangle=\frac{1}{\sqrt{G}}\tilde {\cal W}\left(\frac{1}{\sqrt{G}}\right)=h-1-\frac{T_2}{G}-\frac{T_4}{G(G-1)}\ .
\end{equation}
Namely, from \eqref{hsery} we find  \begin{eqnarray}\nn
    \frac{1}{\sqrt{G}}\frac{\tilde{\cal W}\left(\frac{1}{\sqrt{G}}\right)}{\beta^3}&=&
    \frac{\left(\frac{4 V^3}{\pi }+\frac{2 V}{\pi }\right) \sin ^{-1}(y)^2}{y^2}
      +\left(\frac{4 V^3+2 V}{y^2}+\frac{16 V^3 \sqrt{1-y^2}}{\pi  y^3}\right) \sin ^{-1}(y)
   \\  \label{Wresy}
   &+&\frac{8 V^3
   \left(\sqrt{1-y^2}-1\right)}{y^3}
   +\frac{4V^3}{3 \pi } -\frac{16 V^3}{\pi  y^2}
   -\frac{2 V}{y}-\frac{2 V}{\pi } 
\end{eqnarray}
or equivalently\footnote{here we substituted $\arcsin y+\pi/2=-\arccos y$}
\begin{eqnarray}
\label{Wresy2}
 &&\frac{\tilde{\cal W}(P)}{P\beta^2}=
\frac{ \arccos^2 y-\left(y+\frac{\pi }{2}\right)^2}{\pi   y}
     \\ \nn
     && +
\left(\frac{P^2}{\beta}\right)^2\frac{12 y \arccos^2 y-48 \sqrt{1-y^2} \arccos y+4 y^3-3 \pi ^2 y-48 y-24 \pi }{24 \pi  }
    \end{eqnarray}
where $V$ is defined in \eq{vV}. This formula for the resolvent is one of our main new results. It gives the resummation of graphs with shape of a quadrangulated disc and the expansion of \(\tilde \cW\) in powers of \(G\) generates the moments \(\frac{1}{N}\langle{\rm Tr}(A_4\tilde M)^{2k}\rangle\,,\,\,k=1,2,3,\dots\) in the flattening regime of large area and sparse conical defects scattered over the surface.

An important feature of the result \eqref{Wresy2} is the fact that the whole non-trivial dependence on \(\lambda_c-\lambda,\,\,\beta\)  and \(G\) is hidden (apart from some trivial powers of \(\beta G\)) in the parameter \(y=2\beta  V G\)  which can be viewed as a renormalized boundary cosmological constant.   

The result \eqref{Wresy2}  is rather universal: it should be viewed  as partition function of flat surfaces of disc topology with conical defects    and the boundary represented by a zig-zag line consisting of the links of the boundary of original quadrangulations, as suggested by the image  on figure~\ref{fig:flat}.

Let us now explore some properties of this result. In view of definition of correlators in \eqref{heven} we are interested in the expansion of $h(G)$ in powers of $y$ (or $G$) when $y\to 0$. We find
\begin{eqnarray}
\label{hsery2}
&&
   h-1= \frac{8V^3\beta^3}{y^3}+\beta^2\left(1-\pi\beta\frac{2V^2+1}{4V}\right)\frac{4V^2}{y^2}
   +\frac{2 \beta ^3 V}{y}
   \\ \nn &&
   -\frac{1}{3} \beta ^3 V \left(V^2-1\right) y
      -\frac{2 \beta ^3 V \left(6 V^2-5\right) }{15 \pi }y^2-\frac{1}{20} \beta ^3 V \left(4 V^2-3\right) y^3
   +\dots \  \ .
\end{eqnarray}
Plugging here $y=2\beta VG$ we get
\begin{eqnarray}
\label{hserG}
&&
   h-1= \frac{1}{G^3}+\frac{1}{G^2}\left(1-\pi\beta\frac{2V^2+1}{4V}\right)+\frac{\beta^2}{G}
      \\ \nn
   &&
   -\frac{2}{3} \beta ^4 V^2 \left(V^2-1\right)G
   -\frac{8 \beta ^5 V^3 \left(6 V^2-5\right)}{15 \pi }G^2
   -\frac{2}{5} \beta ^6 V^4 \left(4 V^2-3\right)G^3
   +\dots\ \ .
\end{eqnarray}

From the first few coefficients in the second line we read off, using \eqref{heven}, for instance,
\begin{eqnarray}
    \frac{1}{N}\langle{\rm Tr}(A_4\tilde M)^2\rangle &=& -\frac{2}{3} \beta ^4 V^2 \left(V^2-1\right) + O(\beta^5) \ , \ 
    \\ \nn
    \frac{1}{N}\langle{\rm Tr}(A_4\tilde M)^4\rangle &=& -\frac{8 \beta ^5 V^3 \left(6 V^2-5\right)}{15 \pi } + O(\beta^6) \ , 
    \\ \nn
    \frac{1}{N}\langle{\rm Tr}(A_4\tilde M)^6\rangle &=& -\frac{2}{5} \beta ^6 V^4 \left(4 V^2-3\right)+ O(\beta^7)
\end{eqnarray}
and so on. Notice that while in \eqref{hsery} we know each coefficient of $y^n$ with accuracy of $\sim\beta^3$, when we substitute $y=2\beta VG$ the accuracy of coefficients of $G^n$ will be different because $v\sim \beta$. In \eqref{hserG} we indicated all the information for the coefficients we can extract from \eqref{hsery}\footnote{that means that e.g. the coefficient of $1/G^3$ is given up to $O(\beta)$ corrections, the coefficient of $G$ is up to $O(\beta^2)$ terms, etc}. Below we will describe several nontrivial checks of the computation.

\subsubsection{Checks of the result}

\label{sec:checks}

Let us show that our result \eqref{Wresy2} for the resolvent passes several nontrivial tests. First, the part with negative powers of $G$ in
\eqref{hserG} should match the $T_n$
coefficients as prescribed by \eqref{heven},
which read (recalling that $\epsilon=1$ in our conventions)
\begin{equation}
\label{hpotG}
    h-1=\frac{T_2}{G}+\frac{T_4}{G^2}+\frac{T_4}{G^3}+\frac{T_4}{G^4}+\dots\ \ .
\end{equation}
Since $T_4\simeq 1$ we notice that coefficients of $1/G^n$ with $n\geq 4$ will not be visible in our computation, since when we translate them to $y=G/(2V\beta)$ they will be of order $\beta^n$ and thus lie outside of the $\sim\beta^3$ precision of our result \eqref{hsery}. Indeed, in \eqref{hserG} we see that coefficients of all these terms are zero within our precision. As for the rest, recalling that 
\begin{equation}
    T_{2}={\lambda\beta^2} \ , \ \ T_4=\lambda
\end{equation}
we see that the coefficients of the three terms in the first line of \eqref{hserG} perfectly match the prescribed form \eqref{hpotG}. That serves as a nontrivial consistency check of our result. Notice also that we find no $O(y^0)$ term in the rhs of \eqref{hsery}, which is another check of the result.

As another test, we can compare the $G$ term with the prediction coming from the free energy ${\cal Z}$ that was computed for this model in \cite{Kazakov:1996zm}\footnote{see equation (4.25) there which gives ${\cal Z}$ before taking any limit} and in our regime it is given by equation (5.10) from that paper,
\begin{equation}
\label{Zx}
    {\cal Z}=
    \frac{4t^4}{15\beta^2}(x^6-\frac{5}{2}x^4+\frac{15}{18}x^2-\frac{5}{16}-x(x^2-1)^{5/2})
\end{equation}
where $t=\sqrt{\lambda}\beta^2$ since we have set $\epsilon=1$, 
while we recall that the scaling parameter $x$ from \cite{Kazakov:1996zm} is defined by \eq{defxl}.
Let us note that, curiously, ${\cal Z}$ becomes simply a polynomial (divided by $\beta^2$) when written in terms of $V$ rather than $x$, i.e. plugging $x$ here from \eq{vV} we get
\begin{equation}
    {\cal Z}=\frac{t^4}{\beta^2}\left(\frac{V^6}{15 }-\frac{2 V^4}{15 }+\frac{V^2}{12}\right)\ .
\end{equation}
We see that when expressed in terms of $t$ and $\lambda$ the couplings in \eqref{lambda-beta} become $T_2=\sqrt{\lambda}t$ and $T_{2q}=\lambda$ for $q\geq 2$. This means that  
\begin{equation}
\label{M2cor1}
    \frac{1}{2N}\langle{\rm Tr}(A_4\tilde M)^2\rangle=\d_t{\cal Z}(t,\lambda)
\end{equation}
where we note that one should first write ${\cal Z}$ in terms of precisely $t,\lambda$  variables (excluding $\beta$) before differentiating it. Evaluating the derivative and translating the result to out parameters $V,\beta$ we reproduce the coefficient of the $G$ term in \eqref{hserG},
\begin{equation}
   \frac{1}{N}\langle{\rm Tr}(A_4\tilde M)^2\rangle\simeq  -\frac{2}{3} \beta ^4 V^2 \left(V^2-1\right)\ .
\end{equation}
This is another nontrivial test of our result.

\subsubsection{Pure gravity and almost flat limits}
\label{sec:limits}

In this subsection we will show that our result \eqref{Wresy2} obtained in the flattening limit  interpolates between two interesting critical regimes of this disc partition function : 2d quantum gravity regime and `almost flat' regime.

\paragraph{Pure gravity limit}

The pure gravity limit corresponds to the case when $x\to 1$ as discussed in \cite{Kazakov:1996zm}, and one can see that the free energy \eqref{Zx} has a degree $5/2$ singularity there. Since $x\to 1$ we see from \eqref{defxl} that $\lambda \to \lambda_c$ with the critical value $\lambda_c$ given by \eq{deflc} discussed above. It is also convenient to introduce the `bulk cosmological constant' $\Lambda\propto \lambda-\lambda_c$ via
\begin{equation}
    x-1=\frac{\Lambda}{2}\ .
\end{equation}

Substituting $V$ in terms of $x$ into our result \eqref{Wresy}, we find several important features.
First, plugging $y=G/(2V\beta)$ into \eqref{Wresy} and expanding for $x\to 1$, we find that that
\begin{equation}
\label{Wxm1}
    \tilde \cW=f_1+(x-1)f_2+(x-1)^{3/2}f_3+\dots
\end{equation}
where $f_n$ are some lengthy functions of $\beta$ and $G$. While at first one may expect that the singularity of the resolvent for $x\to 1$ will be of the type $\sim(x-1)^{1/2}$ (for instance, $V$ itself scales as $V\simeq 1/\sqrt{2}-\sqrt{x-1}+\dots$), we see that this is not the case and instead the singularity in \eqref{Wxm1} is  $\sim(x-1)^{3/2}$ due to several nice cancellations.  Thus all correlators $\langle\tr(\tilde M A_4)^n\rangle$ also have a $(x-1)^{3/2}$ behavior. This nontrivial property is in complete agreement with the general prediction of \cite{Kazakov:1985ds, David:1984tx}.

To study the behavior of the disc partition function in this limit, we will need to also send the parameter $y$ (that corresponds to the boundary fugacity) to a critical value, i.e. to a singularity of the resolvent. Notice that the resolvent \eqref{Wresy2} has an apparent branch point at both $y=-1$ and $y=1$, but the first one actually cancels. Thus we will consider $y\to 1$ which we can also parameterise as
\begin{equation}
    y=\sqrt{2}V(1-\zeta)\ .
\end{equation}
Here we introduced, in addition to $\Lambda$, the `boundary cosmological constant' $\zeta$ 
with the normalization  chosen for future convenience. Now, following \cite{Kazakov:1996zm} we consider the scaling limit when $\Lambda,\zeta\to 0,$ keeping fixed the ratio
\begin{equation}
    z=\frac{\sqrt{\Lambda}}{\zeta}
\end{equation}
which implies also $x,y\to 1$. 
Then we find that 
\begin{equation}
    \frac{\tilde{\cal W}}{\beta^3}=\frac{-28-18 \pi +9 \pi ^2}{3 \sqrt{2} \pi }+\frac{\sqrt{2} \left(-12-7 \pi +3 \pi ^2\right) \zeta }{\pi }
    -\frac{8}{3} 
    {\color{blue}\zeta ^{3/2}
   (z-2) \sqrt{z+1}}
   +O\left(\zeta ^2\right)\ .
\end{equation}
The first two terms are regular in $\zeta$ and $z$ and are non-universal. We highlighted in blue the most interesting part, i.e. the 3rd term which is proportional to $\zeta^{3/2}(z-2)\sqrt{z+1}$. This perfectly matches the prediction of \cite{Kazakov:1989bc} for this universal part of the resolvent. That is another highly nontrivial test of our result.

\paragraph{Almost Flat limit}

Another key special case is the almost flat (AF) regime for which the resolvent we are considering was computed in \cite{Kazakov:1995gm}. We can recover that result by setting $V\to 0$ in our formula \eqref{Wresy} for $ \tilde{\cal W}$ (or equivalently keeping only the $1/g$ term in \eqref{Wresy2}). This corresponds to the regime \(x=1+\frac{\sqrt{2}(\lambda_c-\lambda)}{\pi\beta}\gg1\), or \(\beta\ll (\lambda_c-\lambda)\), i.e. the flattening parameter \(\beta\) is much bigger than  the bulk cosmological constant governing the area. In this regime we find
\begin{equation}
\label{Waf1}
    \tilde{\cal W}(P)=\frac{\beta ^2P^2 }{\pi y}((\sin^{-1}y)^2-y^2+\pi(\sin^{-1}y-y))\ .
\end{equation}
Taking into account the difference in notation between our calculation and that of \cite{Kazakov:1995gm}, we perfectly recover equation (4.29) of \cite{Kazakov:1995gm}\footnote{Note a typo in (4.29) of \cite{Kazakov:1995gm}: the r.h.s. should have an extra $t_2$ factor. To compare (4.29) with our result we note that in our case $t_2=\beta^2$ and for $V\to 0$ we see from \eqref{tauv} that $\tau\simeq \frac{\beta}{4V}$. Therefore  the expansion  parameter of \cite{Kazakov:1995gm} defined as $x_{\rm there}=t_2G/(2\tau)$ is actually precisely our $y$. Then it's immediate to see that our result \eqref{Waf1} is the same as (4.29) of \cite{Kazakov:1995gm}.}.

\subsection{Resolvent for ${\rm Tr} M^n$ correlators}

\label{sec:otheres}

Here we will study the resolvent for a different kind of correlators -- namely, the expectation values $\langle{\rm Tr}M^n\rangle$. We recall that it is defined by \eqref{bWR}, 
\begin{equation}
     W(P)
     \equiv
     \frac{1}{N}\langle\frac{1}{P-M}\rangle
\end{equation}
and it can be computed from  $H(h)$ via \eqref{Wpar1} which for convenience we repeat here:
 \begin{equation}
 \label{Whp2}
     W=P-\frac{h}{P}
 \end{equation}
with
 \begin{equation}
 \label{phh2}
     P^2=he^H\ .
 \end{equation}
 
 Notice that for the purely gaussian model we would have $\rho=1$ (corresponding to the empty Young tableau) and thus $H=\log\frac{h}{h-1}$ which gives \begin{equation}
    W=\frac{P-\sqrt{P^2-4}}{2}
 \end{equation}
 which,  as was noticed already in~\cite{Kazakov:1995ae}, corresponds to the tree-like configurations stemming from pairwise Wick contractions of links of the boundary in a planar way. 
We see that $h$ has to stay finite and generic as it is related to the expansion parameter $P$ via \eqref{Whp2}. However, in the `flattening' regime discussed above, $h$ is close to $b$ due to \eqref{hb1}, as the r.h.s. of that equation contains the small parameter $\beta$. Moreover in that regime the cut $[a,b]$ of $H(h)$ collapses and becomes visible as simply a pole, so that inverting the function $H(h)$ (which we need to do in \eqref{phh2}) would only lead to a somewhat trivial  result for the resolvent $W(P)$. This suggests that we have to look for a somewhat different scaling regime here.

 In order to find a suitable scaling limit we will restore the parameter $\epsilon$ that we have set to $1$ above.  While the dependence of the correlators $\langle \tr(A_4\tilde M)^{2n}\rangle$ we computed before on $\epsilon$ is rather trivial (see section \ref{sec:Geps}), this is not the case for the averages $\langle \tr( M)^{2n}\rangle$ which offer the possibility for a more nontrivial behavior. We found that it is natural to still take the `flattening' limit when $\beta,v,\tau$ are all small but at the same time to send now $\epsilon$ to zero together with $\beta$ while keeping their ratio finite,
 \begin{equation}
    D=\beta/\epsilon={\rm fixed} \ , \ \ \beta\to 0\ .
\end{equation}
The parameters $\tau$ and $v$ are taken to be of order $\beta$ like before (see \eqref{tauv}, \eqref{vV}) so that
\begin{equation}
    v=V\beta \ , \ \ \tau\simeq \beta\frac{2V^2+1}{4V}\ .
\end{equation}
We will shortly see that in this regime $h$ naturally stays arbitrary as we wanted, rather than being close to $b$.
The nontrivial scaling also introduces $D$ as an extra tunable parameter in addition to $V$ while keeping the computation similar in many ways to the one done above, and allowing us to use several results from section \ref{sec:exp}. We will be able to compute the resolvent explicitly in this scaling regime which, as we will see, offers a variety of interesting features to explore.  
 
Let us discuss how one can reinstate $\epsilon$ in various expressions. We first notice from \eqref{vtolkb}, \eqref{lameq} that $v$ and $q$ do not have any $\epsilon$-dependence. Then from \eqref{Deltaeq} we see that the dependence of $\Delta$ on $\epsilon$ is simply
\begin{equation}
    \Delta=\frac{\Delta_{\epsilon=1}}{\epsilon^2}
\end{equation}
and moreover
\begin{equation}
     X-1=\frac{(X-1)_{\epsilon=1}}{\epsilon^2}
\end{equation}
with $X=a,b,c,d$. Thus we can use results from section \ref{sec:exppar} such as \eqref{Dexp} and \eqref{brese} which provide these parameters with exponential precision in $\beta$. For instance, at leading order we read off
\begin{equation}
    b\simeq 1+2\beta D^2V\frac{2V^2-3}{3\pi}
\end{equation}
\begin{equation}
    a\simeq 1+4D^2V^2 \ , \ \ \ d\simeq-4D^2/\beta^2 \ , 
\end{equation}
while $b-c$ is exponentially suppressed for small $\beta$. We see an important feature that, unlike in the flattening limit, the cut $[a,b]$ of $H(h)$ remains of finite size in our regime. 

These observations also mean that the expression for $H$, which is the most nontrivial component of the calculation, can be directly extracted from our solution with exponential precision in section \ref{sec:exp} and we do not have to compute any new integrals. We find that
 \begin{equation}
     H=\log\frac{h}{h-b}+2\Phi+i\pi\rho
 \end{equation}
  where $\Phi$ is given by \eqref{Phi}. Expanding it at small $\beta\sim\tau$ we find
    \begin{eqnarray}
    \label{Hep}
&& 
    H=\log \frac{h}{h-b}+[\frac{ \pi  \tau }{2}-\frac{2 \tau  u^2}{\pi }-2 i \tau  u+\frac{2 \beta V (-2+(2 u+i \pi ) \tanh (u))}{\pi }]
   +\dots\ \ .
\end{eqnarray}
Finally,  the relation between $h$ and $u$ reads
\begin{equation}
\label{huep}
    h-b\simeq 4\frac{V^2D^2}{\cosh^2u}
\end{equation}
and we see that in contrast to \eqref{hb1} the r.h.s. does not contain any small parameters and $h$ can be kept arbitrary (with finite $u$ like before) which is just what we would like to have, as explained in the beginning of this section.

The relations we just described are enough to compute the resolvent, namely we express $u$ in terms of $h$ from \eqref{huep}, plug it into \eqref{Hep} and then solve \eqref{phh2} for $h(P)$ perturbatively in $\beta$. The  result to order $O(\beta)$ reads 
\begin{eqnarray}
\label{Wxi2} 
    W&=&\frac{L}{2D  }
    +
    \beta\frac{16D^3}{L^2(4D^2-L^2)}
    \\  \nn
    &\times&
    \left[
    L^2\frac{\sin ^{-1}(\xi )^2-\xi ^2}{4 \pi  \xi  }+\frac{ \xi ^3+12 \sqrt{1-\xi ^2} \sin ^{-1}(\xi )-12 \xi +3 \xi  \sin ^{-1}(\xi
   )^2}{6 \pi  }
   \right]
\end{eqnarray}
where to make the formula more compact we introduced the notation
\begin{equation}
  \xi=V L \ , \ \ \ L=D({P-\sqrt{P^2-4}}).
\end{equation}
The formula~\eqref{Wxi2} represents  another important result of this paper. In the next subsection we will discuss its properties and limiting cases.

The first term in the r.h.s. of \eqref{Wxi2} is the additive  contribution of tree-like configurations of the boundary, as mentioned before. Notice that all the bulk cosmological constant dependence is hidden in the parameter \(\xi\), which can be considered as a renormalized  boundary fugacity controlling its length. The factor \({P-\sqrt{P^2-4}}\) there that comes from summing up the tree-like configurations of parts of the boundary, as well as \(D\) controlling the curvature fluctuations,  also enter that renormalization. Similarly to the parameter \(y\) in~\eqref{Wresy2}, the \(\xi\)  dependence of the two terms in the 2nd line of \eqref{Wxi2} provides the most interesting information about the critical behaviors of disc quadrangulations with this boundary condition. Actually, these two formulas have a very close structure and we will discuss later in this section their similarities and differences.   

\subsubsection{Limits and properties of the result}

The large $P$ expansion of the resolvent \eqref{Wxi2} reads 
\begin{eqnarray}
    && PW=1+\frac{\frac{8 \beta  D^4 V^3 \left(5-6 V^2\right)}{15 \pi }+1}{P^2}+\frac{-\frac{256 \beta  D^6 V^5 \left(10 V^2-7\right)}{315 \pi
   }-\frac{32 \beta  D^4 V^3 \left(6 V^2-5\right)}{15 \pi }+2}{P^4}
   \\ \nn &&
   +\frac{-\frac{512 \beta  D^8 V^7 \left(14 V^2-9\right)}{315 \pi
   }-\frac{512 \beta  D^6 V^5 \left(10 V^2-7\right)}{105 \pi }+\frac{8 \beta  D^4 V^3 \left(5-6 V^2\right)}{\pi }+5}{P^6}+\dots\ \ .
\end{eqnarray}
From the coefficients here we read off the values of the correlators to order $O(\beta)$ as
\begin{eqnarray}
    \frac{1}{N}\langle\tr M^2\rangle &=& 1+\frac{8 \beta  D^4 V^3 \left(5-6 V^2\right)}{15 \pi }
    \\ \nn
    \frac{1}{N}\langle\tr M^4\rangle &=& 2-\frac{256 \beta  D^6 V^5 \left(10 V^2-7\right)}{315 \pi
   }-\frac{32 \beta  D^4 V^3 \left(6 V^2-5\right)}{15 \pi }
\end{eqnarray}
and so on.

One important test is the behavior of the resolvent for $x\to 1$. Like for the case of $\tilde\cW$ discussed in section  \ref{sec:limits}, we expect that the leading singularity should be $(x-1)^{3/2}$ and not $(x-1)^{1/2}$. Indeed, plugging $V$ as a function of $x$ into \eqref{Wxi2} and expanding it, we find
\begin{equation}
    W=g_1+g_2(x-1)+g_3(x-1)^{3/2}+\dots
\end{equation}
where $g_k$ are some functions of $P,D$ and $\beta$. Thus we see that the $(x-1)^{1/2}$ singularity non-trivially cancels and we have instead $(x-1)^{3/2}$ as expected, in perfect agreement with the 2d QG behavior of the sphere partition function with a marked point  \cite{Kazakov:1985ds, David:1984tx}. 

Just like the result in the flattening limit from section \ref{sec:crit2}, our resolvent interpolates between the almost flat and pure gravity regimes. The AF regime corresponds to $V\to 0$ (equivalently, $x\to\infty$) when the resolvent becomes
\begin{equation}
    W=\frac{L}{2D  }
    +
    \beta\frac{4D^3}{4D^2-L^2}
   \frac{\sin ^{-1}(\xi )^2-\xi ^2}{ \pi  \xi  }\ .
\end{equation}
We have also re-checked this result by repeating the derivation of the resolvent starting from the explicit solution given in \cite{Kazakov:1995gm} for the AF case.

Now let us consider the pure gravity limit. Like in section \ref{sec:limits} we expect that it should correspond to sending the parameter $P$ to a branch point of the resolvent. Notice that $W$ has potential branch points at $P=\pm 2$ and also $P=\pm(1/(2VD)+2VD)$ (i.e. $\xi=\pm 1$). As we are interested in the large $P$ expansion (that generates the correlators), the latter two branch points are the relevant ones as they are always closer to infinity. Moreover, similarly to the case of the previous resolvent,  the branch point $P=-1/(2VD)-2VD$ cancels. Thus we will take $P$ to be near the remaining branch point at $P=1/(2VD)+2VD$. Equivalently, we send $L\to \sqrt{2}$. Introducing a convenient normalization, we take the scaling similar to the one in section \ref{sec:limits} 
\begin{equation}
    L=\sqrt{2}(1-\zeta) \ , \ \ \ x=1+\frac{\Lambda}{2} \ , \ \ \ z=\frac{\sqrt{\Lambda}}{\zeta}={\rm fixed}
\end{equation}
where again we view $\Lambda$ as the (renormalized) bulk cosmological constant and $\zeta$ as the boundary one. Then we find for the nontrivial part of  the resolvent the behavior
\begin{eqnarray}
    \frac{1}{\beta}\left[W-\frac{L}{2D}\right]&=&\frac{\left(28-3 \pi ^2\right) D^3}{\pi  \left(3-6 D^2\right)}+\frac{D^3 \left(\left(6 \pi ^2-88\right) D^2-9 \pi ^2+100\right) \zeta
   }{3 \pi  \left(1-2 D^2\right)^2}
   \\ \nn &-&
  \zeta ^{3/2}(z-2)\sqrt{z+1} \frac{4\sqrt{2}   D^3 }{3 \left(2 D^2-1\right)
   }
   +O\left(\zeta
   ^{2}\right)\ .
\end{eqnarray}
We see that the last term  indeed reproduces the correct scaling function
\begin{equation}
    \zeta^{3/2}(z-2)\sqrt{z+1}
\end{equation}
as expected from \cite{Kazakov:1989bc}.

\subsubsection{Comparison of two resolvents}

Curiously, we observed that the two resolvents $\tilde\cW$ and $W$ that we have computed in sections \ref{sec:resflat} and \ref{sec:otheres} are quite closely related to each other. Namely, after a certain simple  substitution and redefinition of the parameters they become equal up to a rational part. The latter may be viewed as not that important since it essentially does not affect the critical behavior, which is determined by branch point singularities.

Concretely, comparing \eqref{Wresy2} and \eqref{Wxi2} we observed that they become almost equal if we identify $y\leftrightarrow\xi$ and also make in the first one the formal replacement $\arccos y\to -\arcsin y$. The meaning of that last replacement is not fully clear to us at the moment, though it looks rather suggestive\footnote{ perhaps it can be viewed as a change of branch in some sense}. To make the matching exact we also need to multiply one of the resolvents by an overall rational factor. As an outcome we find
\begin{eqnarray}
\label{resrel}
&&
    \left(\left.C_1{\tilde{\cal W}\left(P\right)}\right|_{\arccos y\to -\arcsin y,\ y\to \xi}\right)-C_2\left(W-\frac{L}{2D}\right)
    =R_0
\end{eqnarray}
where
\begin{equation}
    C_1=\frac{1}{\beta^3\sqrt{G}} \ , \ \ \ C_2=\frac{4 D^2 L^2-\xi ^4}{2 \beta D^3 L^3} \ , \ \ \ 
    R_0=-\frac{L^3 (\pi  \xi +8)}{\xi ^6}-\frac{L (4 \xi +\pi )}{2 \xi ^3}\ .
\end{equation}
The matching of all the parts including the $\arcsin$ and square root functions between the two resolvents is quite nontrivial and perhaps indicates that the result for the disc partition function shows a certain universality, regardless of how precisely we construct the boundary (as a $\tr (A_4\tilde M)^n$ or a $\tr (\tilde M)^n$ insertion). It would be interesting to understand more rigorously the reason for this matching.

The remaining different rational parts might be attributed to the finite (and generic) size of the boundary in our setup which introduces essentially lattice artefacts. It is worth reminding that  both formulas \eqref{Wresy2}  and  \eqref{Wxi2} represent the disc partition functions of continuous flat manifolds with conical defects scattered over them, but at the same time with the boundary of any finite lattice size, i.e. consisting of any finite number of links. This sounds somewhat eclectic and needs some interpretation. Perhaps one may view such zig-zag boundary as a collection of   conical defects  stuck on it,  making the continuous picture more consistent.   However, at the moment all these possible interpretations are quite speculative and the relation \eq{resrel} is just a nontrivial observation.

\section{Conclusions and prospects}

\label{sec:concl}

The main result of this work is the derivation of disc partition functions  of abstract random 2D manifolds, flat everywhere except conical singularities (with the angle deficits \(\Delta\phi=\pi,0,-\pi,-2\pi,-3\pi,\dots\))      inserted at arbitrary positions and weighted with a certain fugacity controlling the fluctuations of \(|\Delta\phi|\). Since the deficit of angle is akin to the insertion of a curvature defect, such fugacity controls the overall scale of the curvature fluctuations of these manifolds. The area of the manifold and the length of the boundary are weighted with the corresponding cosmological constants.  For two different types of the boundary, we found that these partition functions  are given by expressions \eqref{Wresy2}  and \eqref{Wxi2}. Due to this extra curvature fugacity, the formulas interpolate between the ``almost flat"  regime and the 2D quantum gravity regime. The ``almost flat" regime arises in the limit when the curvature fugacity is such that the overall curvature fluctuations are highly suppressed. It  corresponds to all the negative angle deficit (negative curvature)  concentrated at the boundary of the disc, whereas the positive curvature defects represented by conical singularities with \(\Delta\phi=\pi\) are scattered in the interior of the disc. The partition functions \eqref{Wresy2}  and \eqref{Wxi2} behave slightly differently in this regime, i.e. they depend on the type of the boundary conditions. The 2D quantum gravity regime dominates in the limit opposite to the almost flat case, when the  curvature fugacity favors the proliferation of curvature defects of both signs, i.e. the local metric of the manifold is highly fluctuating. In this regime, both formulas reproduce the well known universal answer for the disc partition function of pure 2d QG. 

It would be also important to compute the sphere and disc partition functions for other types of conical defects via DWG. For example the explicit solution for \(H(h)\) for triangulations, i.e. for the conical defects with angle deficits $\pm\pi$  was already written in Appendix~1  of~\cite{Kazakov:1996zm} in terms of contour integrals (see also \cite{Szabo:1996fj}). However, it is rather involved and needs special efforts for its study.     Whereas in the  2d QG regime the results should be still universal, the limit of almost flat surfaces should  depend on the type of conical defects.

Interestingly, a somewhat similar problem was approached in the series of papers on Jackiw-Teitelboim gravity \cite{Saad:2019lba,Witten:2020wvy,Stanford:2019vob,Turiaci:2020fjj,Maxfield:2020ale}, where one finds the partition functions (with or without boundaries) of the manifolds with constant negative intrinsic curvature \(R<0\) (i.e.  Lobachevski, or \(AdS_2\) metric), also with insertions of any number of conical singularities with arbitrary fixed \(\Delta\phi\). It seems to be tempting to try to reproduce our results \eqref{Wresy2}  and \eqref{Wxi2}   from the results of these papers in the limit \(R\to 0    \), at least in the ``almost flat" limit. However, it appears to be a very singular limit in JT  gravity, so such a comparison is still an open question~\footnote{We thank G. Turiaci for explanations on that issue}.

On the other hand, one of our motivations for the current research was to construct and solve  a DWG  model that would imitate directly the sum over 2D manifolds with dominating \(AdS_2\) background. At first sight, it seems to be easy: for example we can take in the DWG model~\eqref{ZDWGh} the following choice of the constants:  \(t_q^*=\delta_{q,4}\) (quadrangulations) and \(t_q=g\delta_{q,4}+\gamma\delta_{q,6}\) (only zero or negative curvatures). This guarantees the \(AdS_2\) background, at least on average, for large quadrangulations. However, the Euler theorem tells us that this can be achieved only with a boundary, i.e. for the disc, and all the positive curvature should be concentrated at this boundary. It is precisely the last condition which is difficult to achieve in DWG model. For both  types of the boundary conditions, \eqref{Wresy2}  and \eqref{Wxi2}  contain both positive and negative curvatures in the bulk. It seems that to guarantee such a localisation of positive curvature on the boundary we have to compute a different disc partition function. One possibility is to take  \(\mathbb{W}_L=\left<\frac{1}{N}\tr(A_4^{-1}\tilde M)^L\right>\), with the same condition for quadrangulations \(\frac{1}{N}\tr A_4^q=\delta_{q,4}\). That means that near the boundary, instead of the the squares we will have the triangles contributing the angles \(\pi/3\). For example, two neighboring squares and one triangle give at such vertex the positive angle deficit \(2\pi-(\pi/2+\pi/2+\pi/3)=2\pi/3\). Consequently, we should be able to localize all such defects at the boundary. It would be fascinating to explore this direction further. 

It would be also interesting to investigate in-depth the analytic structure of the disc partition function of quadrangulations in the exponential approximation~\eqref{Grese}, \eqref{hue2}. It might contain new interesting critical regimes, inaccessible in the ``flattening" approximation used in~\eqref{Wresy2}  and \eqref{Wxi2}.       

Another future direction is to properly formulate and explore the spectral curve for the generic DWG model. It would be very interesting to uncover what type of integrability arises in such matrix models, and also whether and how some kind of tau functions  could appear here (even at finite $N$).  

We hope to return to these questions in the future.

\section*{Acknowledgements}
 
We thank \"O.~G\"urdo\(\check{\rm g}\)an for help with simplifying the integrals in section \ref{sec:integrals}. We are also grateful to B.~Basso, I.~Kostov, A.~Mironov, D.~Serban, G.~Turiaci for discussions and comments.

\appendix

 \section{Derivation of the partition function and details on characters}
\label{CharExp_sec} 

Here we discuss some technical details for the derivations in sections \ref{sec:dwgch} and \ref{sec:characters}, based on \cite{Kazakov:1995ae}.

Using the completeness relation for characters, the integral in \eq{DWGMM}  can be expanded into characters of couplings \(t_q\) and \(t_q^*\) as follows:
\begin{align}
\label{chichi}
 Z(t,t^*)&=\sum _{h}\chi_{\{h\}}[t]~\int d^{N^2}M\,\exp N \tr \left(-\frac{N}{2}M^2\right)\notag\chi_{\{h\}}(AM)
\\
&=\sum _{h}\frac{1}{d_h}\chi_{\{h\}}[t]~\chi_{\{h\}}[t^*]
\int {\cal D}M_i\Delta(M)\det_{k,l}(M_k^{h_l})\exp N \tr \left(-\frac{N}{2}M^2\right)
\end{align}
where  \(M_i\) are eigenvalues of \(M\). The integral then can be computed explicitly and is equal to \begin{equation}\label{Mint}
\int {\cal D}M_i\Delta(M)\det_{k,l}(M_k^{h_l})\exp N \tr \left(-\frac{N}{2}M^2\right)
=c(N)d_h\frac{\prod_i(h^e_i-1)!!h^o_i!!}{
\prod_{i,j}(h^e_i-h^o_j)}~\ .
\end{equation}

 The Schur characters with all but one vanishing couplings can be computed explicitly.  For the character with the couplings  \(t_q=s \delta_{q,m}\) we have (see Appendix~8 of \cite{Kazakov:1995ae}):
\begin{align}\label{Schur_m}\chi_{\{h\}}(s)=
c \quad (s)^{\frac{1}{m}\sum_j h_j}  \,\,\left({N\over m}\right)^{{1\over m}\Sigma_ih_i}\quad
\prod_{\epsilon=0}^{m-1}{\Delta(h^{(\epsilon)})\over
\prod_i\bigl({h^{(\epsilon)}_i-\epsilon\over m}\bigr)!}
\,\,{\rm sgn}\bigl[\prod_{0\leq\epsilon_1<\epsilon_2\leq(m-1)}
\prod_{i,j}(h^{(\epsilon_2)}_i-h^{(\epsilon_1)}_j)\bigr]\end{align}
where \(h^{(\eps)}=\eps\mod{m}\).

For our principal case of interest in this paper, \(m=4\),  we have from here
\begin{align}\label{Schur4}\chi_{\{h\}}(A_4)\sim  \,\,\quad
\prod_{\epsilon=0}^{3}{\Delta(h^{(\epsilon)})\over
\prod_i\bigl({h^{(\epsilon)}_i-\epsilon\over m}\bigr)!}
\,\,{\rm sgn}\bigl[\prod_{0\leq\epsilon_1<\epsilon_2\leq3}
\prod_{i,j}(h^{(\epsilon_2)}_i-h^{(\epsilon_1)}_j)\bigr]\end{align}
where \(h^{(\eps)}=\eps\mod{4}\).

In order to compute the correlator $\langle \tr M^{2L}\rangle$ as discussed in section \ref{sec:characters}, we use the formula similar to \eqref{tchi=chi}  
\begin{equation}\label{t=chi}
\tr M^L\,\cdot\chi_{\{h\}}[M]=\sum_{j=1}^{N}\chi_{\{h+L\delta_j\}}[M]\,\qquad  \end{equation}
 and, performing the gaussian integral over \(M\) via  \eqref{Mint} we obtain at the saddle point of  \eqref{chichi} the result \eq{Mcorr} given in the main text.

\section{Mathematica notation for elliptic functions}
\label{app:ell}

Here we summarize the Mathematica notation corresponding to some elliptic functions we use in the text.

\beq
\begin{array}{r|l}
    \text{Function}& \text{Mathematica command}
    \\
    \hline
    K & \text{EllipticK}[k^2]
    \\
    K' & \text{EllipticK}[1-k^2]
        \\
    v={\rm sn}^{-1}\left(\sqrt{\frac{a-c}{a-d}},k'\right) & \text{InverseJacobiSN}[\sqrt{\frac{a-c}{a-d}},1-k^2]
 \\
 E(v,k') & 
 {\rm EllipticE}[{\rm ArcSin}[{\rm JacobiSN}[v,1-k^2]]]
\end{array}
\eeq

\section{Details on the exponential approximation}

\label{app:exp}

Let us show that, as discussed in section \ref{sec:integrals}, with exponential precision in $\tau$ we can do the following simplifications in the integral in r.h.s. of \eq{intba}:
\begin{itemize}
    \item Replace the upper integration limit by $\infty$ 
    \item Use the approximation \eq{rhoe} for $\rho$
    \item Use \eq{dhe} for the quantity $J(w)\equiv \frac{dh}{dw}\frac{1}{h(u)-h(w)}$
\end{itemize}

To demonstrate the 1st point, we recall that $K\simeq \pi/(2\tau)$ with exponential precision, and for $w$ greater than this (large) value we can estimate the integrand as
\beq
\rho\sim 1, \ \ J(w) \sim e^{-2w}
\eeq
where the second estimate follows from \eq{dhe}.
This gives for the part we are dropping
\beq
    \int_K^\infty dw J(w)\rho(w)\sim e^{-\pi/\tau}
\eeq
so it is indeed exponentially small.

Concerning the 2nd point, notice that \eq{rhoe} holds as long as $q'^2\sinh 2u\ll 1$, i.e. up to values $u\sim 1/\tau$. For larger $u$ again we can replace $\rho\to 1$ in order to estimate the integral, and by the same argument as above we see that the region where \eq{rhoe} is not applicable gives an exponentially small contribution.

Finally to demonstrate the 3rd point, we compute the quantity $J(w)$ by starting from \eq{defu}, writing it is a series in $k$ around $k=1$ and then looking at the large $w$ behavior. While at leading order we get \eq{dhe}, at higher orders we find potentially dangerous terms, e.g. at 2nd order we get a term that at large $w$ behaves as $\sim (k^2-1)^2 e^{2w}$. Since the integral goes from $0$ to $K$ we can estimate its contribution as $\sim (k^2-1)^2e^{2K}$. Although the second factor here is large and of order $e^{2K}\sim e^{\pi/\tau}$, it is still outmatched by the first one since  $(k^2-1)^2\sim e^{-2\pi/\tau}$ so in combination they give $\sim e^{-\pi/\tau}$, i.e. an exponentially small contribution again. One can repeat a similar estimate at higher orders and verify explicitly that any extra contributions are exponentially suppressed.

\bibliographystyle{unsrt}
\bibliography{AdS2_biblio}

\end{document}